\documentclass[12pt]{article}

\usepackage{amsmath,epsfig,ulem}

\newcommand{\sll}{{\tilde{l}}}
\newcommand{\slR}{{\tilde{l}_{\rm R}}}
\newcommand{\slL}{{\tilde{l}_{\rm L}}}
\newcommand{\msl}[1]{m_{\tilde{l_{#1}}}}
\newcommand{\smu}{{\tilde{\mu}}}
\newcommand{\smuR}{{\tilde{\mu}_{\rm R}}}
\newcommand{\smuL}{{\tilde{\mu}_{\rm L}}}
\newcommand{\se}{{\tilde{e}}}
\newcommand{\seR}{{\tilde{e}_{\rm R}}}
\newcommand{\seL}{{\tilde{e}_{\rm L}}}

\newcommand{\stR}{{\tilde{\tau}_1}}

\newcommand{\sn}{{\tilde{\nu}}}
\newcommand{\sne}{{\tilde{\nu}_e}}
\newcommand{\snl}{{\tilde{\nu}_l}}

\newcommand{\msn}[1]{m_{\tilde{\nu}_{#1}}}

\newcommand{\mseL}{m_{\tilde{e}_{\rm L}}}

\newcommand{\cha}{\tilde{\chi}}
\newcommand{\neu}{\tilde{\chi}^0}
\newcommand{\mcha}[1]{m_{\tilde{\chi}^\pm_{#1}}}
\newcommand{\mneu}[1]{m_{\tilde{\chi}^0_{#1}}}

\newcommand{\hh}[1][1]{\tfrac{#1}{2}}

\newcommand{\dMW}{\delta \MW^2}

\def\mathswitch#1{\relax\ifmmode#1\else$#1$\fi}
\def\mathswitchr#1{\relax\ifmmode{\mathrm{#1}}\else$\mathrm{#1}$\fi}
\newcommand{\PW}{\mathswitchr W}
\newcommand{\PZ}{\mathswitchr Z}

\newcommand{\Pe}{\mathswitchr e}

\newcommand{\MW}{\mathswitch {M_\PW}}
\newcommand{\MZ}{\mathswitch {M_\PZ}}

\newcommand{\me}{\mathswitch {m_\Pe}}

\newcommand{\mf}{m_f}
\newcommand{\scrs}{{}}
\newcommand{\sw}{\mathswitch {s_{\scrs\PW}}}
\newcommand{\cw}{\mathswitch {c_{\scrs\PW}}}

\newcommand{\gev}{\,\, \mathrm{GeV}}
\newcommand{\mev}{\,\, \mathrm{MeV}}
\newcommand{\re}{\Re e \,}

\newcommand{\SLASH}[2]{\makebox[#2ex][l]{$#1$}/}

\newcommand{\Eslash}{\SLASH{E}{.5}\,}
\newcommand{\RR}{{\rm R}}
\newcommand{\LL}{{\rm L}}
\newcommand{\eR}{e_{\rm R}}
\newcommand{\eL}{e_{\rm L}}

\newcommand{\anc}{\rule{0mm}{0mm}}
\newcommand{\lesim}{\,\raisebox{-.1ex}{$_{\textstyle <}\atop^{\textstyle\sim}$}\,}

\newcommand{\drbar}{{\mathswitch {\overline{\rm DR}}} }
\newcommand{\OO}{{\mathcal O}}

\newcommand{\mycaption}[1]{\caption{\sl #1}}

\begin{document}
\thispagestyle{empty}

\def\thefootnote{\fnsymbol{footnote}}

\begin{flushright}
DESY 04--133\\
FERMILAB--Pub--04/156--T\\
\end{flushright}

\vspace{1cm}

\begin{center}

{\Large\sc {\bf Sneutrino Production 
at \boldmath{$e^+e^-$} Linear Colliders:\\[1ex] [Addendum to Slepton Production]}}
\\[3.5em]
{\large
{\sc
A.~Freitas$^{1}$%
,
A.~von Manteuffel$^{2}$
and
P.~M.~Zerwas$^{2}$
}
}

\vspace*{1cm}

{\sl
$^1$ Fermi National Accelerator Laboratory, Batavia, IL 60510-500, USA

\vspace*{0.4cm}

$^2$ Deutsches Elektronen-Synchrotron DESY, D--22603 Hamburg, Germany
}

\end{center}

\vspace*{2.5cm}

\begin{abstract}

Complementing the preceding study of charged scalar leptons, the sector of the
neutral scalar leptons, sneutrinos, is investigated in a high-precision
analysis for future $e^+e^-$ linear colliders. 
The theoretical predictions for the cross-sections are
calculated at the thresholds for non-zero widths and in the continuum 
including higher-order
corrections at the one-loop level.
Methods for measuring the sneutrino masses and the 
electron-sneutrino-gaugino Yukawa couplings are presented,
addressing theoretical problems specific for the sneutrino channels.

\end{abstract}

\def\thefootnote{\arabic{footnote}}
\setcounter{page}{0}
\setcounter{footnote}{0}

\newpage


\section{Introduction}

The precision analysis of charged scalar leptons (sleptons) at future
high-energy $e^+e^-$ and $e^-e^-$ colliders has been studied for the first
two generations in a preceding report \cite{slep}, establishing precise
theoretical calculations which include non-zero width and rescattering
effects at the thresholds and the complete set of one-loop corrections
in the continuum. The corresponding process of stau production, including
mixing effects, has been discussed in Ref.~\cite{stau}.

In this addendum the work will be extended to the neutral slepton sector,
{\it i.e.} the production of sneutrinos in $e^+e^-$ annihilation, 
\begin{equation}
e^+e^- \to \snl\snl^* \quad \mbox{for}\quad l = e,\mu,\tau. 
\end{equation}
Note that the $\snl$ are assumed to
be the partners of the left-handed sneutrinos, since if the right-handed
neutrino masses are close to the GUT scale, the R-sneutrino masses are
expected to be so heavy \cite{blair:00p} that mixing with the L-sneutrinos
is irrelevant.

We will elaborate in particular the onset of the electron-sneutrino
($\sne$) cross-section near threshold, which allows a determination of the
$\sne$ mass. The accuracy of the mass measurement through threshold scans
is superior to other methods using decay spectra of the sneutrinos
produced in the continuum \cite{nauenberg}. Belonging to the same
iso-doublet, the masses of the sneutrinos and the L-sleptons are related
by a sum rule involving the D term of the supersymmetry (SUSY) Lagrangian.
This symmetry relation can be tested stringently.

While sneutrino production of the second and
third generation in $e^+e^-$ collisions is mediated by s-channel $Z$-boson
exchange, $\sne$ pair production in addition involves t-channel chargino
exchange, thereby being sensitive to the electron-sneutrino-chargino 
$e^\mp\sne\cha^\pm$ Yukawa
couplings. From the measurement of $\sne$ production one can therefore
scrutinize the identity of the gauge and Yukawa couplings of the electroweak
SU(2) sector.

To provide precise theoretical descriptions for sneutrino production, we
have calculated the production cross-sections near threshold including
non-zero width and off-shell effects, while in the continuum the 
complete one-loop corrections for on-shell sneutrino production
are presented. The phenomenological analyses are based on the Minimal
Supersymmetric Standard Model (MSSM), adopting as an example the Snowmass
Point SPS1a \cite{sps}. Effects of initial-state beamstrahlung as well as
decays of the sneutrinos are taken into account. The final results
demonstrate that measurements of the $\sne$ sneutrino masses and of the
$e\sne\tilde{W}$ couplings are possible at the per-cent level,
supplementing the similarly precise picture of the charged slepton sector
\cite{slep}.

In Section \ref{basics}  the general
features of sneutrino production are summarized at leading order. Section
\ref{threshold} presents sneutrino pair production at threshold and analyses
the expected accuracies in threshold scans. In Section \ref{continuum} the
pair production of sneutrinos in the continuum is analyzed at the one-loop
level, and applied for the  determination of SUSY Yukawa couplings.
Conclusions are summarized in Section~\ref{concl}.


\section{\hspace{-2mm}Basics of Sneutrino Production and Decay}
\label{basics}

For definiteness, the analysis is restricted to the Minimal Supersymmetric
Standard Model (MSSM), including sneutrinos with chiral quantum
number L, while R-sneutrinos are assumed absent as argued before.
We adopt the conventions and notations introduced in Ref.~\cite{slep}.

\subsection{Production Mechanisms}

\begin{figure}[t]
\vspace{2em}
\centering
\begin{tabular}{cc}
\raisebox{-2mm}{\epsfig{file=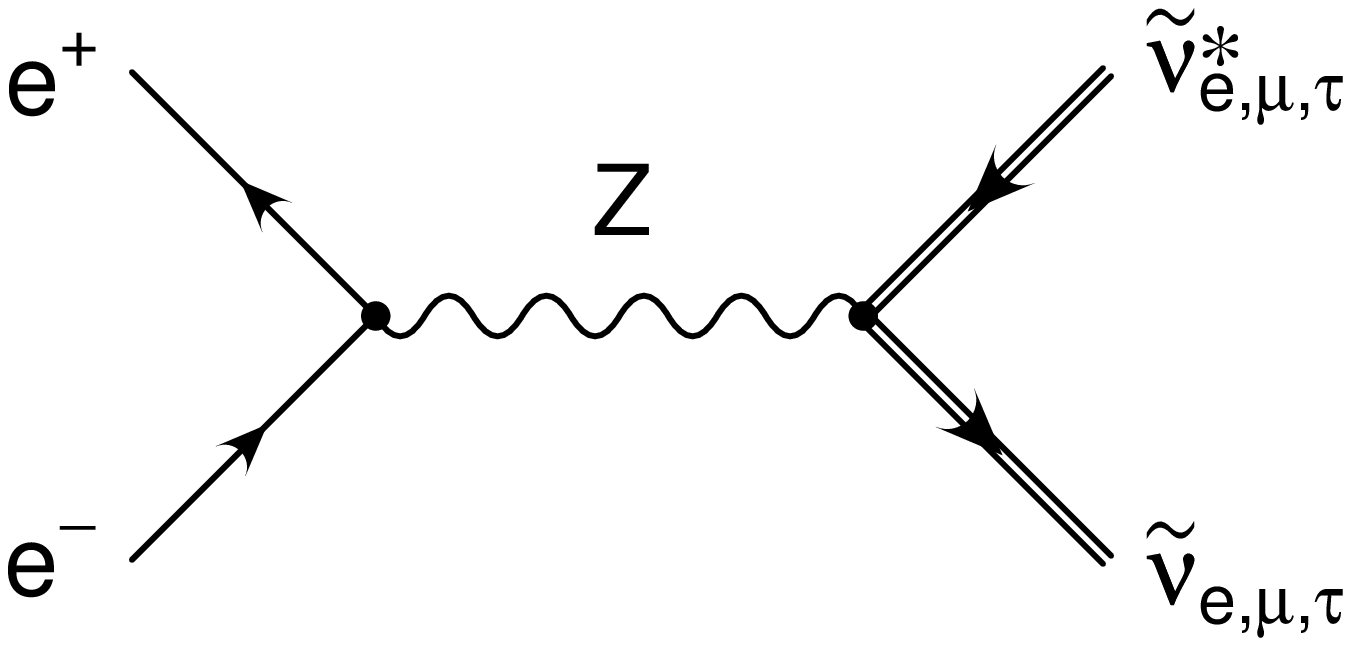,width=8cm}} &
\epsfig{file=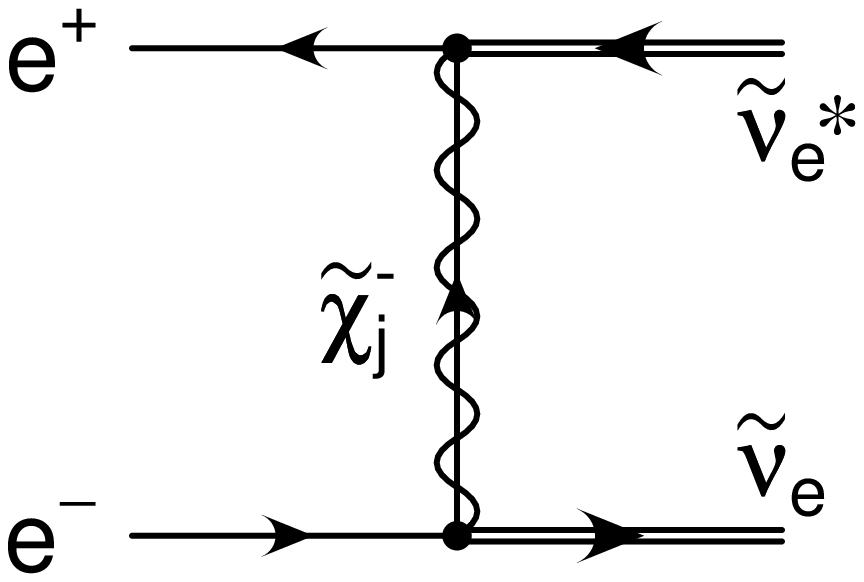,width=4.5cm} \\[-1em]
 (a)$\qquad\quad$ & $\quad$(b)
\end{tabular}
\mycaption{Generic leading-order diagrams for the pair production 
of sneutrinos in $e^+e^-$ annihilation.}
\label{fig:cdiag}
\end{figure}

Muon- and tau-sneutrinos $\sn_{\mu,\tau}$ are generated in diagonal pairs 
via s-channel $Z$-boson exchange in $e^+e^-$
collisions, see Fig.~\ref{fig:cdiag}~(a). Electron-sneutrino
$\sne$ pair production proceeds in addition through
chargino $\cha^\pm_j$ [$j=1\dots 2$] exchanges in the t-channel, cf.
Fig.~\ref{fig:cdiag}~(a,b).
Since only sneutrinos in with chiral index L are produced, 
the conservation of chiral quantum number in the interactions demands 
sneutrinos to be produced in P-wave states, for both s- and t-channel 
mechanisms near threshold. This leads to the characteristic
$\beta^3$ onset of the excitation curves near threshold, with $\beta$ denoting
the velocity of the final-state sneutrinos.

The Born cross-sections for sneutrino production by polarized beams read
\begin{align}
\sigma[\eR^+ \, \eL^- \to \snl \, \snl^*] &=
  \frac{2 \pi \alpha^2}{3s} \beta^3 \,
  \frac{(1- 2\sw^2)^2}{16 \sw^4 \cw^4} \biggl ( \frac{s}{s-\MZ^2} \biggr )^2
  && [l = \mu,\tau], \label{eq:bornbegin}
\displaybreak[0] \\[2ex]
\sigma[\eR^+ \, \eL^- \to \sne \, \sne^*] &=
  \frac{2 \pi \alpha^2}{3s} \beta^3 \,
  \frac{(1- 2\sw^2)^2}{16 \sw^4 \cw^4} \biggl ( \frac{s}{s-\MZ^2} \biggr )^2
\nonumber \\
&\;+ \frac{\pi \alpha^2}{\sw^4 s}
 \sum_{j=1}^2 \sum_{k=1}^2 |V_{j1}|^2 \, |V_{k1}|^2 \, h^{jk} \label{eq:seBorn}
 \\
&\;+ \frac{2 \pi \alpha^2}{\sw^2 s}
 \sum_{j=1}^2 |V_{j1}|^2  \frac{1-2\sw^2}{4\sw^2\cw^2} \, \frac{s}{s-\MZ^2} \, f^j
 , \nonumber
\displaybreak[0] \\[1ex]
\sigma[\eL^+ \, \eR^- \to \snl \, \snl^*] &=
  \frac{2 \pi \alpha^2}{3s} \beta^3
  \frac{1}{4 \cw^4} \biggl ( \frac{s}{s-\MZ^2} \biggr )^2
 && [l = e,\mu,\tau], \label{eq:bornend}
\displaybreak[0] \\[2ex]
\sigma[\eR^+ \, \eR^- \to \snl \, \snl^*] &=
        \sigma[\eL^+ \, \eL^- \to \snl \, \snl^*] = 0
 && [l = e,\mu,\tau],\nonumber
\end{align}
with
\begin{align}
f^j &= \Delta_j \beta - \frac{\Delta_j^2 - \beta^2}{2} \,
        \ln \frac{\Delta_j + \beta}{\Delta_j - \beta}, \displaybreak[0] \\[1ex]
h^{jk} &= \left\{
\begin{array}{ll}
  \displaystyle -2 \beta + \Delta_j \ln\frac{\Delta_j + \beta}{\Delta_j - \beta}
    \quad & j = k \\
  \displaystyle \frac{f^k - f^j}{\Delta_j - \Delta_k}
    & j \neq k
\end{array} \right. ,
\end{align}
where 
\begin{equation}
\Delta_j = 2 (\msn{l}^2 - \mcha{j}^2) /s -1
\quad {\rm and} \quad
\beta = \sqrt{1 - 4 \msn{l}^2/s}, \label{eq:kindia}
\end{equation}
and $V_{ij}$ is the mixing matrix for positively charged charginos (see
Ref.~\cite{slep}). The electromagnetic coupling $\alpha$ should be taken 
at the scale $Q^2=s$.

\begin{figure}[tb]
\epsfig{file=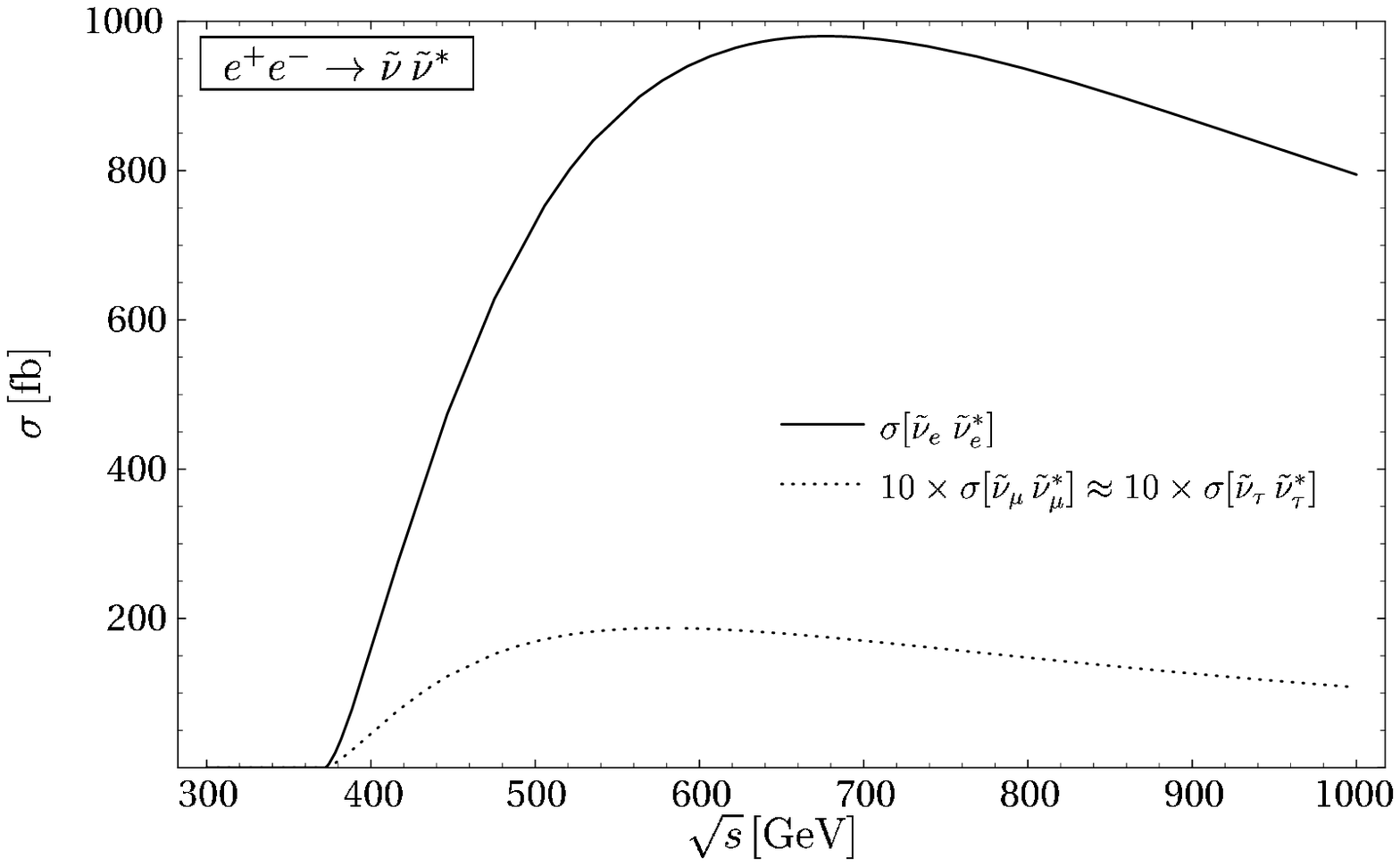,width=6in}
\vspace{-1ex}
\mycaption{Born cross-sections for sneutrino pair production in unpolarized
$e^+e^-$ annihilation. The cross-section for $\sn_\tau$ production
is very similar to the $\sn_\mu$ production cross-section.
The mass values in the SPS1a scenario are collected in Tab.~\ref{tab:sps1}.
}
\label{fig:snXsec}
\end{figure}

In the t-channel chargino exchange amplitudes, the contribution of the higgsino
component of the chargino states can be neglected owing to the  small
electron-sneutrino-higgsino coupling proportional to the electron mass. The
exchange of relatively light  charginos with dominant gaugino component  leads
in general to electron-sneutrino production cross-sections that are more than
an order of magnitude larger than for the other two flavors. Such a scenario
is realized in the reference point SPS1a \cite{sps}, as shown in
Fig.~\ref{fig:snXsec}.

The angular distributions for the sneutrino production processes with
polarized beams read
\begin{align}
\frac{{\rm d}\sigma}{{\rm d}\Omega}[\eR^+ \, \eL^- \to \snl \, \snl^*] &=
  \frac{\alpha^2}{4s} \beta^3 \sin^2 \theta \,
  \frac{(1- 2\sw^2)^2}{16 \sw^4 \cw^4} \biggl ( \frac{s}{s-\MZ^2} \biggr )^2
  \qquad\qquad\qquad [l = \mu,\tau],
\displaybreak[0] \\[2ex]
\frac{{\rm d}\sigma}{{\rm d}\Omega}[\eR^+ \, \eL^- \to \sne \, \sne^*] &=
  \frac{\alpha^2}{4s} \beta^3 \sin^2 \theta \,
  \frac{(1- 2\sw^2)^2}{16 \sw^4 \cw^4} \biggl ( \frac{s}{s-\MZ^2} \biggr )^2
\nonumber \\
+ \frac{\alpha^2}{4\sw^4 s} \beta^3 &
 \sum_{j=1}^2 \sum_{k=1}^2 |V_{j1}|^2 \, |V_{k1}|^2 \, 
 \frac{\sin^2 \theta}{\bigl[\Delta_j - \beta\cos\theta\bigr]
 			\bigl[\Delta_k - \beta\cos\theta\bigr]}
 \\
+ \frac{\alpha^2}{2\sw^2 s} \beta^3 &
 \sum_{j=1}^2 |V_{j1}|^2  \frac{1-2\sw^2}{4\sw^2\cw^2} \, \frac{s}{s-\MZ^2} \,
 \frac{\sin^2 \theta}{\Delta_j - \beta\cos\theta}, \nonumber
\displaybreak[0] \\[1ex]
\frac{{\rm d}\sigma}{{\rm d}\Omega}[\eL^+ \, \eR^- \to \snl \, \snl^*] &=
  \frac{\alpha^2}{4s} \beta^3 \sin^2 \theta
  \frac{1}{4 \cw^4} \biggl ( \frac{s}{s-\MZ^2} \biggr )^2
  \qquad\qquad\qquad\qquad\quad [l = e,\mu,\tau],
\end{align}
with $\theta$ being the angle between the incoming $e^-$ and the outgoing
$\sn$ particles.
For muon- and tau-sneutrino production, the angular sparticle distributions
follow the $\sin^2\theta$ rule typical for P-wave production. While near the
threshold the angular distribution of electron-sneutrino production is also
$\propto\sin^2\theta$, the t-channel chargino exchange amplitude peaks in the
forward region near $\cos\theta \approx 1$ for high center-of-mass energies:
\begin{equation}
\frac{{\rm d}\sigma}{{\rm d}\cos\theta}[e^+ \, e^- \to \sne \, \sne^*]
 \propto \sum_{j,k}
 \frac{1-\cos^2 \theta}{\bigl[\Delta_j - \beta
 \cos\theta\bigr] \bigl[\Delta_k - \beta
 \cos\theta\bigr]}
 \stackrel{s \gg \msn{}^2}{\longrightarrow}
 \frac{1+\cos\theta}{1-\cos\theta} ,
\end{equation}
as expected from helicity analysis and t-channel exchange.

\subsection{Decay Mechanisms}
\label{decays}

The sneutrinos are typically expected to decay into light neutralino or
chargino states with large gaugino components. The leading-order decay widths
for these two-particle decays are given by
\begin{align}
\Gamma[\snl \to \nu_l \, \neu_j] &= \alpha \, |X_j|^2 \; \msn{l}
 \Biggl ( 1-\frac{\mneu{j}^2}{\msn{l}^2} \Biggr )^{\!\!2}
\qquad [i = \LL/\RR, \; j = 1\dots4], \label{eq:neudec} \\
\Gamma[\snl \to l^- \, \cha^+_k] &= \frac{\alpha}{4} |V_{k1}|^2 \, \msn{l}
 \Biggl ( 1-\frac{\mcha{k}^2}{\msn{l}^2} \Biggr )^{\!\!2}
\qquad [k = 1,2], \label{eq:chadec}
\end{align}
where 
$X_j$ and $V_{kl}$ account for the neutralino and chargino mixings \cite{slep}.

\renewcommand{\arraystretch}{1.2}
\begin{table}[tb]
\begin{center}
\begin{tabular}{|c||l|r@{\:}ll|}
\hline
Sparticle & Mass $m$ [GeV] & \multicolumn{3}{c|}{Decay modes} \\
 & Width $\Gamma$ [GeV] & & & \\
\hline \hline
$\slR = \seR/\smuR$ & $m = 142.72$ & $\slR^-$ & $\to l^- \, \neu_1$ & 100\% \\
        &       $\Gamma = 0.21$ & & & \\
\hline
$\slL = \seL/\smuL$ & $m = 202.32$ & $\slL^-$ & $\to l^- \, \neu_1$ & 48\% \\
        &       $\Gamma = 0.25$ && $\to l^- \, \neu_2$ & 19\% \\
        &                       && $\to \nu_l \, \cha^-_1$ & 33\% \\
\hline
$\snl = \sne/\sn_\mu$ & $m = 185.99$ & $\snl$ & $\to \nu_l \, \neu_1$ & 87\% \\
        &       $\Gamma = 0.16$ & & $\to \nu_l \, \neu_2$ & 4\% \\
        &                       & & $\to l^- \, \cha^+_1$ & 10\% \\
\hline
$\sn_\tau$ & $m = 185.05$ & $\sn_\tau$ & $\to \nu_\tau \, \neu_1$ & 89\% \\
        &       $\Gamma = 0.15$ && $\to \nu_\tau \, \neu_2$ & 3\% \\
        &                       && $\to \tau^- \, \cha^+_1$ & 8\% \\
\hline \hline
$\neu_1$ & $m = 96.18$ & \multicolumn{2}{c}{---} & \\
\hline
$\neu_2$ & $m = 176.62$ & $\neu_2$ & $\to \seR^\pm \, e^\mp$ & 6\% \\
        &       $\Gamma = 0.020$ && $\to \smuR^\pm \, \mu^\mp$ & 6\% \\
        &                       && $\to \stR^\pm \, \tau^\mp$ & 88\% \\
\hline
$\cha_1^\pm$ & $m = 176.06$ & $\cha^+_1$ & $\to \stR^+ \, \nu_\tau$ & 100\% \\
        &       $\Gamma = 0.014$ &&  & \\
\hline
\end{tabular}
\end{center}
\vspace{-1em}
\mycaption{Masses, widths and main branching ratios of sleptons and of the 
light neutralino and chargino states at Born level
for the reference point SPS1a \cite{sps,spsval}.}
\label{tab:sps1}
\end{table}
Masses, widths and branching ratios for the reference point SPS1a
\cite{sps} are collected
in Tab.~\ref{tab:sps1}. 
In the SPS1a scenario the decays into the
lightest chargino and the second lightest neutralino, both of which are
predominantly wino, are suppressed due to the small mass differences.
As a consequence, the dominant channel is the decay into
the bino-like lightest neutralino, leading to a completely invisible final
state.

To obtain visible sneutrino-pair signals, we will therefore focus on the
channel with one sneutrino decaying invisibly directly to the lightest
neutralino, while the other sneutrino decays into a chargino and a charged
lepton. Due to the relatively large value of $\tan\beta = 10$ in SPS1a and
the large stau mixing, charginos decay almost completely into a $\tau$
final state. As a result, the physical signal for sneutrino production in
this channel consists of a charged lepton with the sneutrino flavor
$l=e,\mu,\tau$, one (additional) $\tau$ with opposite charge, and missing
energy, $e^+e^- \to l^\pm \tau^\mp + \Eslash$. For $\sne$ and $\sn_\mu$
production this leads to a clear detector signature with an overall
branching ratio of 17\%. However, the analysis for $\sn_\tau$ production
is experimentally very demanding, since only a pair of tau leptons is
generated in the final state, with large backgrounds from two-photon
interactions and other supersymmetric processes such as direct stau pair
production.


\section{Threshold Production and Mass Measurements}
\label{threshold}

As a consequence of chiral quantum number and angular momentum conservation,
all sneutrino pairs are produced in $e^+e^-$ annihilation near threshold in
P-waves for both the s-channel and t-channel exchange mechanisms.
This leads to the characteristic $\beta^3$ behavior of the excitation
curve as a function of the velocity $\beta$ of the produced particles.
Since the produced sparticles are neutral, no Coulomb rescattering effects
enhance the cross-sections at the thresholds.

When studying non-zero width effects, the
off-shell production of the sneutrinos requires the analysis of the complete
resonance decays and the corresponding continuum backgrounds with the same 
final states, {\it i.e.} the process
\begin{equation}
        e^+e^- \to l^\pm \! \stackrel{_{_{(-)}}}{\nu_l} \neu_1 \, \cha^\mp_1,
\end{equation}
including all Feynman diagrams leading
to this final state.

The theoretical and phenomenological analysis of the threshold
cross-sections for sneutrino production makes use of the methods set up in
Ref.~\cite{thr1,susy02,slep} for charged sleptons. As pointed out already
in Ref.~\cite{Miz01}, due to the small production cross-section for muon-
and tau-sneutrino production together with the possibly suppressed
branching ratio into the final state $l^\pm \!\!
\stackrel{_{_{(-)}}}{\nu_l} \!\! \neu_1 \, \cha^\mp_1$, the expected event
rates near threshold are too low to perform a mass measurement from a
threshold scan. For electron-sneutrino production, on the other hand, the
signal rates are sufficiently high to allow a statistically significant
threshold mass measurement. Subtracting the background cross-section by
extrapolation from energies below threshold, the onset of the sneutrino
excitation curve is in general sharp enough to determine precisely the
sneutrino mass independent of the absolute normalization of the signal
cross-section. This can be realized operationally by performing a
multi-parameter fit including the overall normalization as a free
variable. Within this frame the threshold production will be analyzed in
the following for electron-sneutrinos.


\begin{figure}[tb]
(a) \underline{Double resonance diagram} \hspace{2ex}
(b) \underline{Single resonance diagrams}
\vspace{0.3cm} \\
\phantom{} \epsfig{file=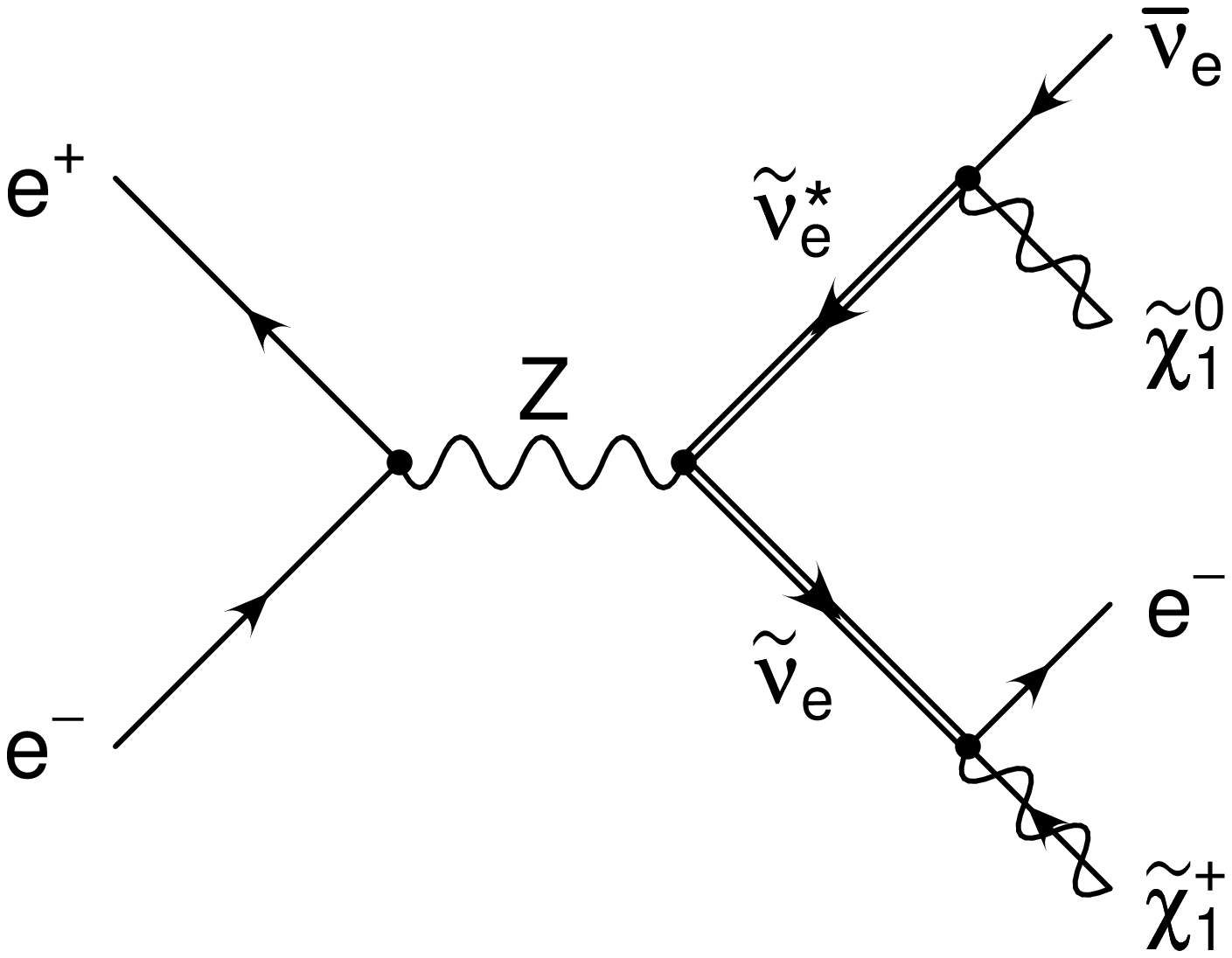,width=5cm, bb=113 240 540 544}
\hspace{5ex}
\epsfig{file=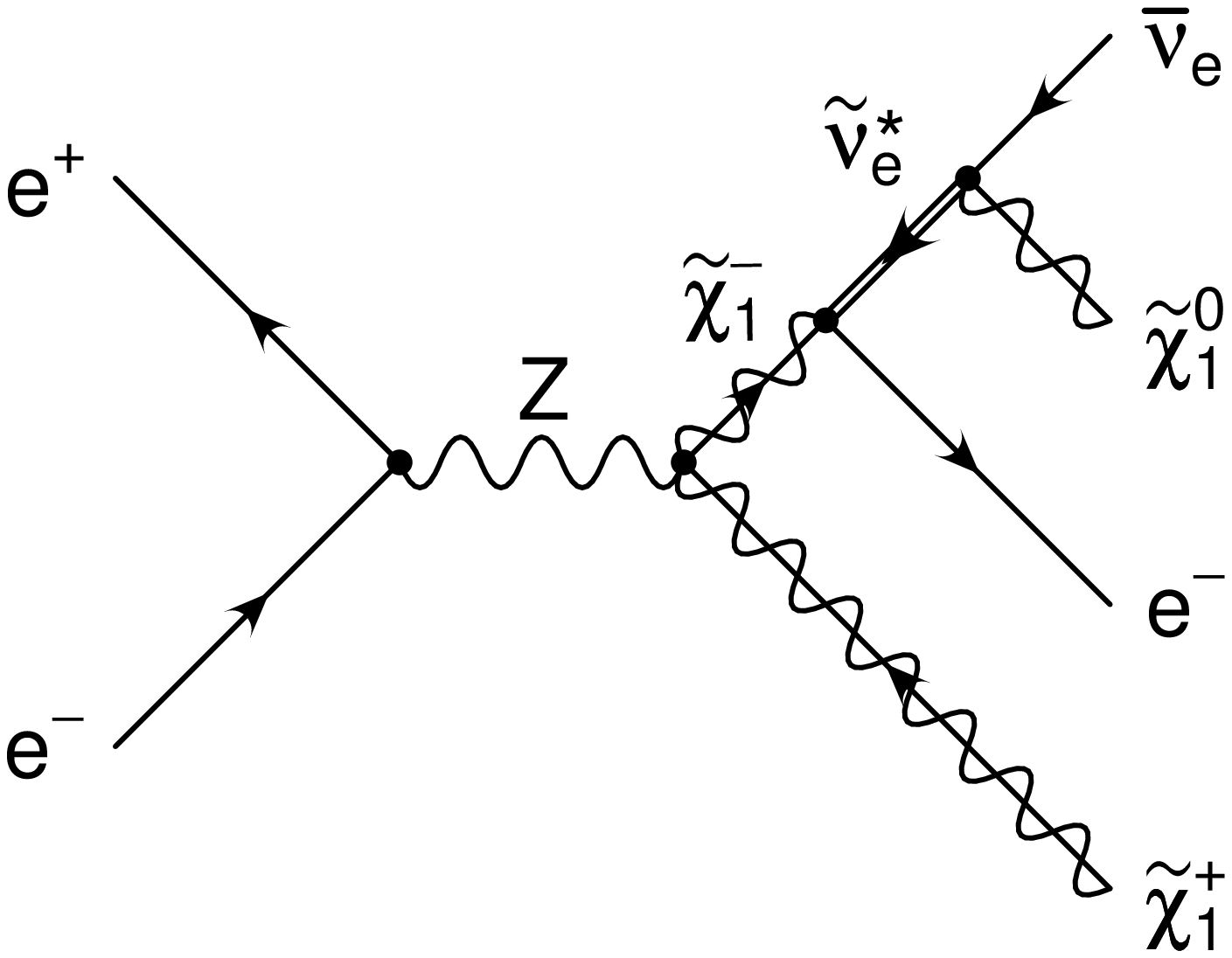,width=5cm, bb=113 240 540 544}
\epsfig{file=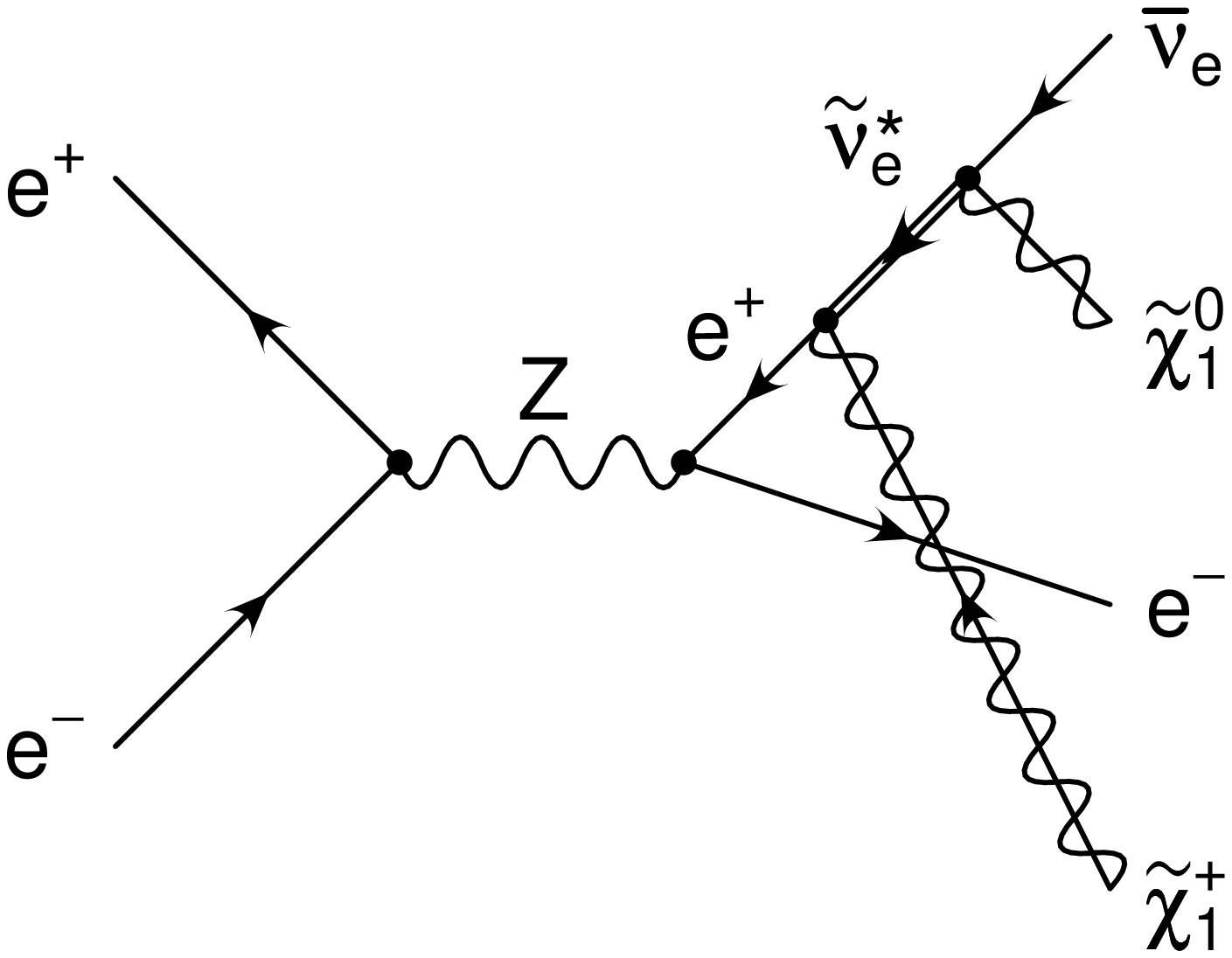,width=5cm, bb=113 240 540 544} \\[3ex]
\phantom{}\hspace{5cm} \hspace{5ex}
\epsfig{file=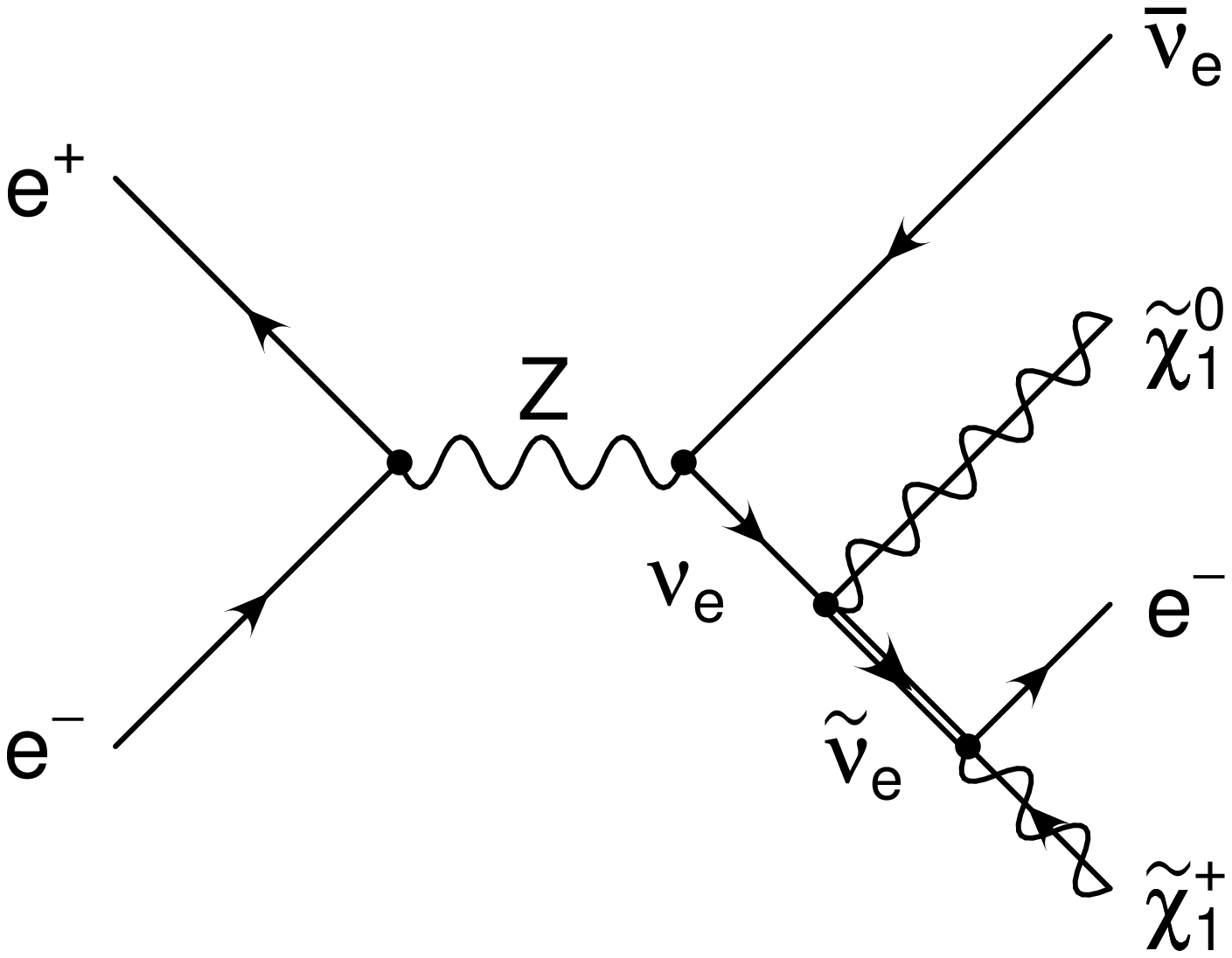,width=5cm, bb=113 240 540 544}
\epsfig{file=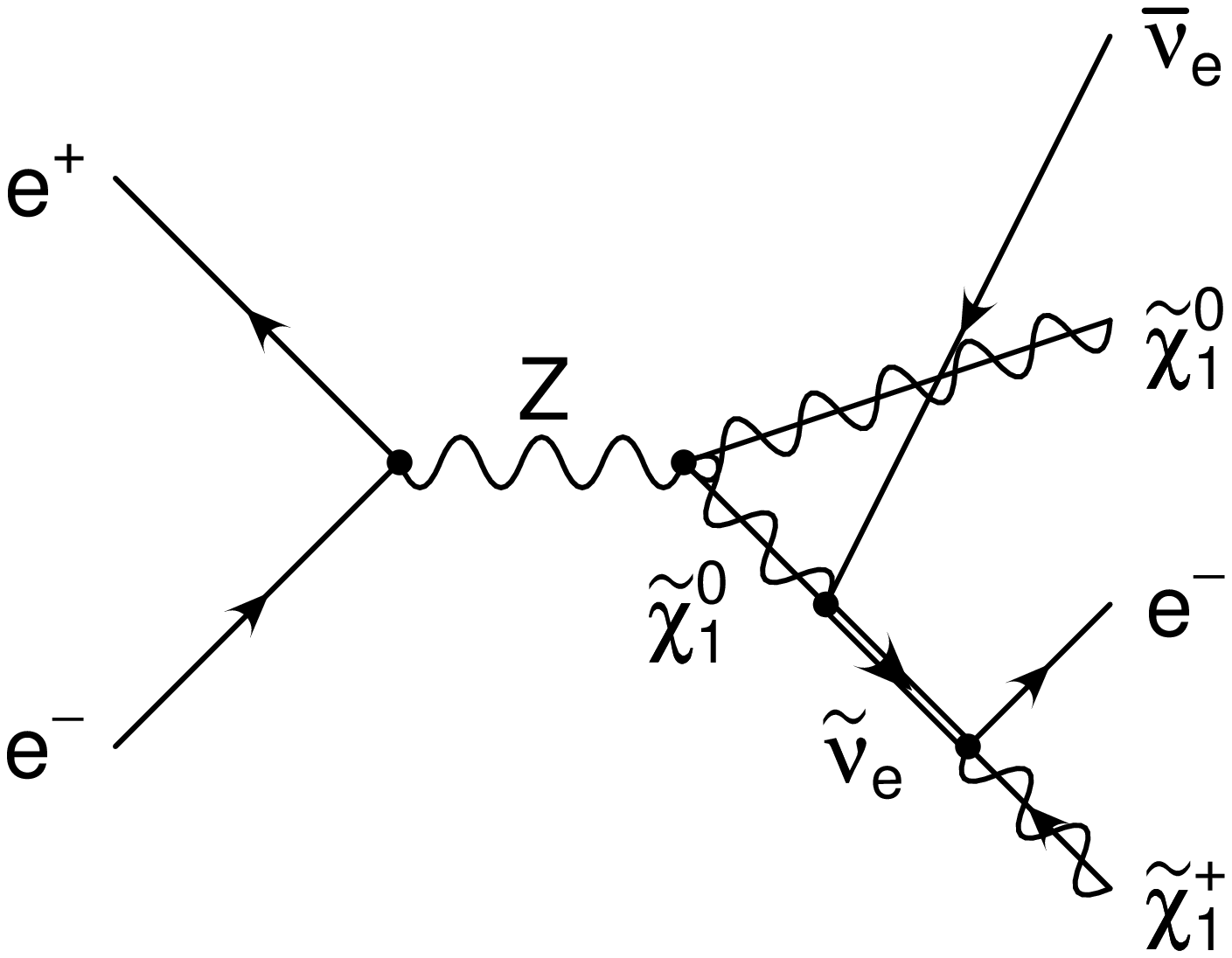,width=5cm, bb=113 240 540 544}
\mycaption{The doubly and singly resonant contributions
to the process $e^+e^- \to e^- \bar{\nu}_e \, \neu_1 \, \cha^+_1$.}
\label{fig:signal}
\end{figure}
The leading contribution to the $e^\pm \!\! \stackrel{_{_{(-)}}}{\nu_e} \!\!
\neu_1 \, \cha^\mp_1$ final states is given by the double resonance diagram
shown in Fig.~\ref{fig:signal}~(a). The non-zero sneutrino width $\Gamma_{\sn}$
is incorporated by shifting the mass in the sneutrino propagators into the
complex plane, $m_{\sn}^2 \;\to\; M_{\sn}^2 = m_{\sn}^2 -
im_{\smu}\Gamma_{\sn}.$  To keep the amplitude gauge invariant,
the double-resonance diagram of Fig.~\ref{fig:signal}~(a) must be supplemented
by the single-resonance diagrams of Fig.~\ref{fig:signal}~(b). In addition to
the s-channel $Z$-exchange diagrams shown in the figure, there exist
corresponding diagrams with t-channel chargino exchange.

\label{fsana}

Besides the signal sneutrino channels,
the final state in the general process $e^+e^- \to e^\pm \tau^\mp + \Eslash$
receives backgrounds from a large variety of other processes.
Within the SUSY sector itself, the main background arises from
selectron pair production $\seR^\pm \seL^\mp$ and $\seL^+ \seL^-$, followed by
the decays $\se_{\LL,\RR}^\pm \to e^\pm \neu_1$ and $\seL^\mp \to \nu_e
\cha^\mp_1 \to \nu_e \nu_\tau \tau^\mp \neu_1$.
To a lesser extent, also
pair production of charginos or neutralinos
with subsequent (cascade) decays contributes. 

In addition, pure Standard Model (SM) processes, in particular $W$-boson pair
production ($W^+W^-$) and single $W$-boson production ($W^\pm e^\mp \nu_e$),
leading to the final state $e^\pm \tau^\mp \nu_e \nu_\tau$, have to be
taken into account. They are generically large and need to be reduced by
appropriate cuts. Since the missing energy signature in the Standard Model
background is generated by two neutrinos, which can be considered massless,
but the sneutrino signal involves two additional massive neutralinos, a cut
on the missing energy proves to be effective. In the SPS1a scenario, the
sneutrino signal near threshold is concentrated at visible energies of $E_{\rm
vis} \leq 0.15 \times \sqrt{s} \approx 60$ GeV, whereas the main background
leads to much large values of the visible energy. Therefore the requirement
$
E_{\rm vis} \leq 0.15 \times \sqrt{s} \label{eq:ecut}
$
reduces the background by about an order of magnitude, while leaving the
sneutrino signal intact. In addition a few general cuts are applied
to parametrize the acceptance regions of the detector and to reduce 
the background from soft events.
The explicit values for the cuts are summarized in Tab.~\ref{tab:WZ1}.%
\renewcommand{\arraystretch}{1.3}%
\begin{table}[tb]
\centering
\begin{tabular}{|p{6.2cm}|p{5.3cm}|l|}
\hline
Condition & Variable & Accepted range \\
\hline \hline
Reject leptons in forward/\newline backward region 
from Bhabha/\newline M\o ller scattering
& lepton polar angle $\theta_{
\rm l}$ &
  $|\cos \theta_{\rm l}| < 0.95$ \\
Reject soft leptons/jets from \newline radiative photon splitting and
$\gamma$-$\gamma$ background
 & lepton/jet energy $E_{\rm l}$ & $E_{\rm l} > 5$ GeV 
\\
Reject missing momentum in forward/backward region
from particles lost in the beam pipe &
  missing momentum polar \newline angle $\theta_{\vec{p}_{\rm miss}}$ &
  $|\cos \theta_{\vec{p}_{\rm miss}}| < 0.90$ \\
Angular separation between \newline $e$ and $\tau$ lepton &
  angle $\phi_{\rm e\tau}$ between electron \newline and tau
  jet & $|1- \cos \phi_{\rm e\tau}| > 0.015$ \\
\hline
\hline
Reject events with large visible energy, mainly from SM background &
  visible energy $E_{\rm vis} = E_e + E_\tau$ &
  $E_{\rm vis} \leq 0.15 \times \sqrt{s}$ \\
\hline
\end{tabular}
\mycaption{Cuts to reduce the main Standard Model and SUSY backgrounds and
to account for the detector geometry and resolution.}
\label{tab:WZ1}
\end{table}
\renewcommand{\arraystretch}{1}%
Additional acollinearity cuts \cite{CDRMartyn,slep}, would not improve the
signal-to-background ratio.

The signal rates can be further enhanced by using beam polarization. As evident
from \mbox{(\ref{eq:bornbegin}--\ref{eq:bornend})}, the optimal polarization
combination for $\sne$ production is a left-polarized $e^-$ and a
right-polarized $e^+$ beam; with realistic polarization degrees of 80\% 
for electrons and
50\% for positrons \cite{pola}.

For this setup, we have calculated the threshold cross-section for
electron-sneutrino production including the backgrounds from 
supersymmetric and Standard Model sources. The calculation is supplemented by
initial-state radiation (ISR) and beamstrahlung effects.%
\begin{figure}[tb]
\epsfig{file=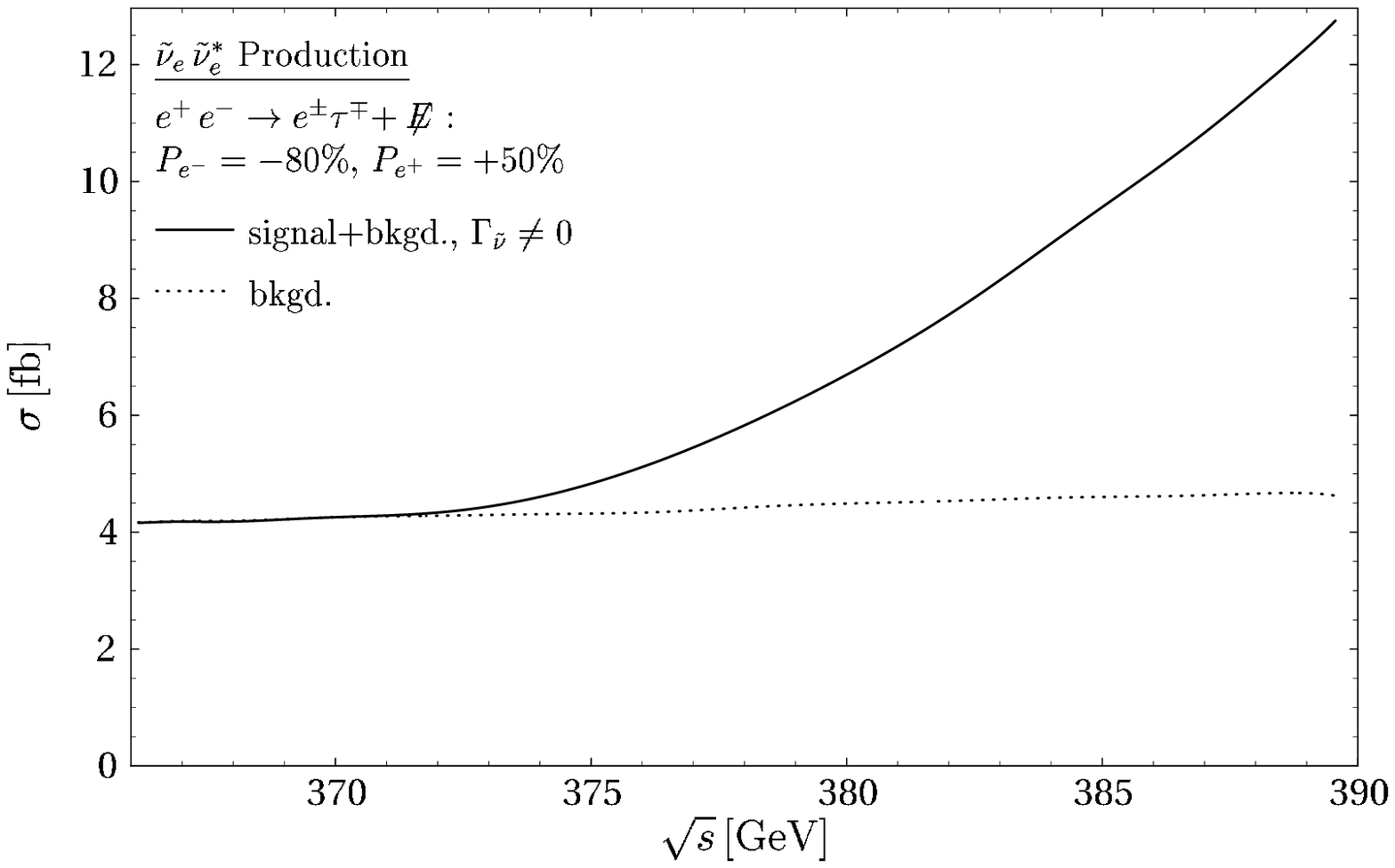,width=6in}
\mycaption{The excitation curves for $\sne$ pair production over Standard Model and
supersymmetric backgrounds for $e^+e^-$ annihilation.
The signal is enhanced with beam polarization as indicated, where
$(+)$ corresponds to right-handed and $(-)$ to left-handed polarization.
}
\label{fig:thr}
\end{figure}
The resulting excitation curve is depicted in Fig.~\ref{fig:thr},
showing the final prediction for the signal with the backgrounds
added on.

We estimate the precision for the sneutrino mass measurement 
from a threshold scan based on
data simulated at 5 equidistant points in a center-of-mass energy range of
20 GeV in the threshold region for $\sne$ pair production, assuming a
luminosity of
10 fb$^{-1}$ per scan point. Since the
backgrounds are sufficiently flat, they can be approximated by a straight
line. Therefore the excitation curve can be fitted in a model-independent way by
using four free parameters: the sneutrino mass and width, a constant
scale factor for the absolute normalization
of the excitation curve and a constant background
level. The last two parameters render the fit independent on the details of the
supersymmetric model, in particular the {\it a priori} unknown branching
fractions of the sneutrinos.
For the reconstruction of the mass a binned likelihood method is employed, where
the likelihood variable is a function of the four parameters mentioned above.
The fit result for the mass determination is finally
\begin{equation}
  \msn{e} = 186.0^{+1.1}_{-0.9} \gev. \label{eq:thrres}
\end{equation}
This demonstrates that the mass measurement at threshold is feasible
with a precision of better than
1\% and in fact surpasses the expected precision from decay energy
spectra \cite{nauenberg}. The result is not affected by the $\sne$ width in the
theoretically expected range.
For the decay width of the sneutrino only a wide upper
bound can be set, $\Gamma_{\sne} \lesim 0.9 \gev.$


\section{Continuum Production and Yukawa Couplings} 
\label{continuum}

The precision analysis of sneutrino production in the continuum is used on the
one hand
to determine the sneutrino and chargino masses from the energy spectra of the
decay products, a method complementary to the threshold scans \cite{nauenberg}.

In addition, it can be used to test the fundamental structure of the
supersymmetric theory, the identity of the lepton-sneutrino-gaugino ($l 
\snl \tilde{W}$) Yukawa
couplings with the associated lepton-neutrino-gauge boson ($l\nu_l W$)  
couplings of the first
generation. The cross-section for $\sne$ pair production is sensitive to the
Yukawa couplings through the t-channel chargino exchange.
This section will address the theoretical techniques necessary for controlling
the higher-order corrections and, subsequently, the phenomenological 
evaluation of the
sneutrino cross-section measurement for the purpose of Yukawa coupling
extraction.

\subsection{Structure of one-loop corrections}

For the general calculational techniques, the computer algebra tools and the
regularization and renormalization of the MSSM at the one-loop level, the
reader is referred to Ref.~\cite{slep}. Here the framework will be extended to
illuminate the specific points in  the renormalization of sneutrinos. As
before, the renormalization of masses and physical fields are performed in the
on-shell scheme, {\it i.e.} the particle masses are identified with the
physical propagator poles, and the on-shell physical fields are normalized to
unity.

Since L-sneutrinos and charged L-sleptons are grouped together in an SU(2)
doublet, the renormalization of neutral and charged sleptons must be treated
conjointly. In the limit of vanishing lepton masses, the charged L- and
R-sleptons do not mix, so that the R-slepton can be separated from the
sneutrino and L-slepton states. 

Since SU(2) invariance allows only for one soft supersymmetry breaking 
parameter
$m_{\tilde{L}_{\LL}}$ for
the sneutrino and the charged L-slepton of one generation, the two 
masses
are interdependent. Based on the
tree-level relations between particle masses, soft-breaking terms and $D$-terms
\begin{align}
\msl{\LL}^2 &= m_{\tilde{L}_{\LL}}^2 + \MZ^2 \, \cos 2\beta \, (-\hh + \sw^2) +
{\cal O}(\me^2), \\
\msn{l}^2 &= m_{\tilde{L}_{\LL}}^2 + \hh \MZ^2 \, \cos 2\beta,
\end{align}
the L-sneutrino and L-slepton masses within each generation are related at 
Born
level according to
\begin{equation}
\msn{l}^2 - \msl{\LL}^2 = \MW^2 \cos 2\beta. \hspace{2.5cm} \mbox{[Born level]}
\label{eq:slborn}
\end{equation}
This relation is modified by higher-order corrections. Adopting the on-shell
renormalization scheme for the slepton masses, the mass difference is 
shifted by a
finite, but non-zero amount,
\begin{align}
\msn{l}^2 - \msl{\LL}^2 &\to \MW^2 \cos 2\beta + \Delta\msn{l}^2, \\
\Delta\msn{l}^2 &= \re \!\! \left \{ 
\Sigma^{\sll_\LL}(\msl{\LL}^2) - \Sigma^{\snl}(\msn{l}^2) \right \}
+ \cos 2\beta \; \dMW 
  - \MW^2 \sin^2 2\beta \; \frac{\delta \tan\beta}{\tan\beta}. \label{eq:snshift}
\end{align}
$\hat{\Sigma}^{\sll_\LL, \snl}$ are the unrenormalized self-energies of the
L-slepton and the sneutrino, respectively. Their UV-divergence is canceled by
the counterterms for the $W$ mass and the Higgs mixing term $\cos 2\beta$.
Here the $W$ mass is renormalized on-shell, while \drbar renormalization is employed
for the parameter $\tan\beta$ \cite{slep}.

By {\it fiat}, the physical on-shell masses of the charged sleptons are
identified with the Born-level masses. As a result, the sneutrino masses are
shifted by the finite amount $\Delta\msn{l}^2 \equiv \msn{l}^2 - m^2_{\snl,\rm
Born}$ from the corresponding Born value to the physical on-shell value through
the one-loop corrections in eq.~\eqref{eq:snshift}. 
The sneutrino mass shift is exemplified in Tab.~\ref{tab:olmass} for the
reference point SPS1a.
\renewcommand{\arraystretch}{1.2}
\begin{table}[tb]
\begin{center}
\begin{tabular}{|l||c|c|c|}
\hline
 & $\slR = \seR/\smuR$ & $\slL = \seL/\smuL$ & $\snl = \sne/\sn_\mu$ \\
\hline &&&\\[-2.5ex]
\begin{minipage}[c]{2cm} tree-level \\ one-loop \end{minipage} &
  142.72 GeV & 202.32 GeV &
  \begin{minipage}[c]{2.2cm} 185.99 GeV \\ 186.34 GeV \end{minipage} \\[1.5ex]
\hline
\end{tabular}
\end{center}
\vspace{-1em}
\mycaption{Tree-level and one-loop corrected masses of the first and second
generation sleptons for the SPS1a reference point \cite{sps}.
The charged slepton masses are defined identical at
tree and loop level
in the on-shell scheme adopted here; the sneutrino mass is shifted
correspondingly.}
\label{tab:olmass}
\end{table}

In the calculation of the loop contributions it is necessary to use tree-level
masses throughout in order to preserve gauge invariance\footnote{This feature
has been
checked explicitly by using a general covariant $R_\xi$ gauge and verifying the
gauge-parameter independence of the total result.}.
For the phase-space integration, on the other hand, the matrix elements need to
be expressed in terms of the physical masses for the external
final-state sneutrinos.
Both requirements can be technically realized by a systematic 
expansion of the complete loop-corrected matrix element. Defining the matrix
element ${\cal M}(\msn{}^2,\msn{}^2)$ as a function of the physical
masses of the two produced sneutrinos and expanding it around
the Born sneutrino  masses yields 
\begin{equation} 
{\cal M}(\msn{}^2,\msn{}^2) ={\cal M}(m^2_{\sn,\rm Born},m^2_{\sn,\rm Born})
  + \Delta \msn{}^2 \left [ \frac{\partial}{\partial \msn{}^2}
  {\cal M}(\msn{}^2,\msn{}^2) \right ]_{\msn{}^2=m^2_{\sn,\rm Born}}
  \!\!+ \OO(\Delta \msn{}^4) .
\end{equation}
Expanding this expression
up to next-to-leading order in perturbation theory and observing that 
$\Delta \msn{}^2$ represents already a one-loop contribution then leads to
\begin{equation} 
\begin{aligned}
{\cal M}(\msn{}^2,\msn{}^2) &=
 {\cal M}_{(0)}(m^2_{\sn,\rm Born},m^2_{\sn,\rm Born})
 + {\cal M}_{(1)}(m^2_{\sn,\rm Born},m^2_{\sn,\rm Born})\\
  &+ \Delta \msn{}^2 \left [ \frac{\partial}{\partial \msn{}^2}
  {\cal M}_{(0)}(\msn{}^2,\msn{}^2)
  \right ]_{\msn{}^2=m^2_{\sn,\rm Born}}
  + \mbox{higher orders} ,
\end{aligned}  \label{eq:snexp}
\end{equation}
where the indices in parentheses indicate the loop order. As evident from
\eqref{eq:snexp}, the next-to-leading corrections receive two contributions:
the one-loop corrections to the sneutrino production matrix element, ${\cal
M}_{(1)}$, with Born masses in all loop expressions, and a contribution
originating from the one-loop mass shift of the sneutrinos.

Note that the shift $\Delta \msn{}^2$, beyond its role in the present
theoretical context, can in principle be experimentally accessed directly
through precision measurements of the sneutrino and selectron masses. With the
expected precision of $\msn{e}$ and given the per-mill accuracy of $\mseL$, the
measured difference of the physical masses $\msn{e} - \mseL$ starts to become
sensitive to the quantum correction $\Delta \msn{}^2$.

A finite mass shift similar to \eqref{eq:snshift} arises for all scalar
partners of the left-handed Standard Model fermions, {\it i.e.} also for the
squark sector. It may be noted that while these sfermion mass
shifts do affect the one-loop calculation of sfermion production, they need not
be taken into account for the one-loop corrections to sfermion decays presented
in Ref.~\cite{Guasch:01}, since the Born matrix elements for sfermion decays do
not depend on the mass of the decaying sfermion.

In many facets,
the radiative corrections to sneutrino production are characteristic for
sparticle production processes. In particular, supersymmetric loop
contributions can lead to corrections that do not decouple for large
superpartner masses \cite{eYuk, superoblique}. While in general  quantum
effects are reduced with increasing mass of the virtual particles, 
broken symmetries giving rise to mass splittings within particle
multiplets elude this argument. Accordingly, the breaking of supersymmetry can
generate corrections that grow logarithmically with the mass splitting between
SM particles and their SUSY partners.

In slepton production, non-decoupling SUSY effects can arise in the loop
corrections to lepton-slepton-gaugino Yukawa vertices, leading to
logarithmically enhanced effective Yukawa couplings \cite{superoblique2,slep},
while the gauge boson vertices are protected by gauge invariance. Therefore only
the production matrix elements of first-generation sneutrinos, {\it i.e.}
$\sne\sne^*$ production, involving t-channel chargino exchange, are affected by
non-decoupling corrections from other SUSY particles. Potentially large
corrections can arise in particular from virtual quark and squark loops for
very large squark masses, so that sneutrino cross-sections can be sensitive
to high squark mass scales \cite{Nojiri:97}.

In addition to the matrix elements of the production cross-sections,
non-decoupling SUSY corrections typically also occur in the loop corrections to
the mass matrices of the superpartners. These corrections are reflected in
shifts to mass relations, such as the sneutrino mass shift $\Delta \msn{}$ in 
\eqref{eq:snshift}.

Another remarkable property is the appearance of anomalous threshold
singularities within vertex and box diagrams \cite{slep}, which show up as
discontinuities in the cross-section as a function of the center-of-mass
energy. Anomalous thresholds correspond to configurations in a vertex loop
correction, for instance, where all particle lines become on-shell
simultaneously, which is easily kinematically realizable  in SUSY theories 
as a consequence of the diversified sparticle spectrum \cite{slep}.

\subsection{Production Cross-Sections}

The complete virtual $\OO(\alpha)$ corrections, as outlined above, are
renormalized using the on-shell
definition for all masses, while the electromagnetic coupling $\alpha$ is
evaluated at the scale of the center-of-mass energy $Q = \sqrt{s}$, so that the
large logarithmic corrections $\propto \log s/\mf^2$ from light fermion loops
in the running of $\alpha(Q^2)$ are absorbed automatically.

\begin{figure}[tp]
{\centering
\psfig{figure=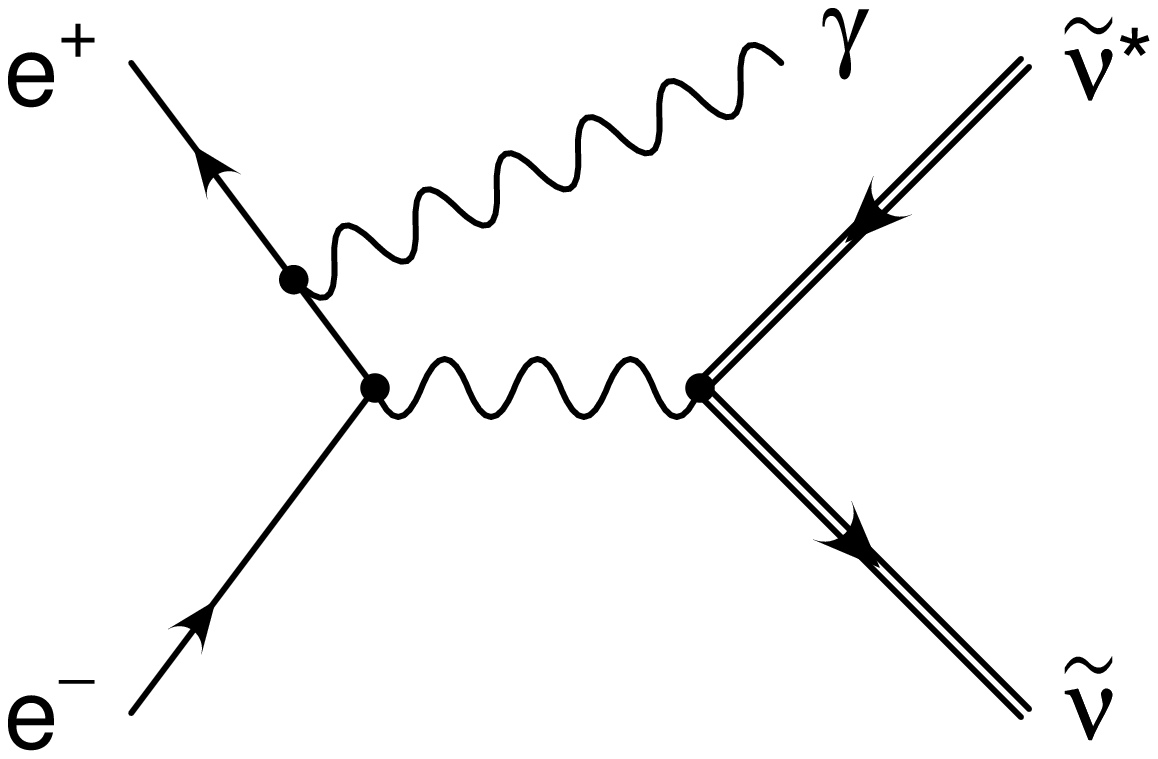, width=5.3cm}
\hspace{10mm}
\psfig{figure=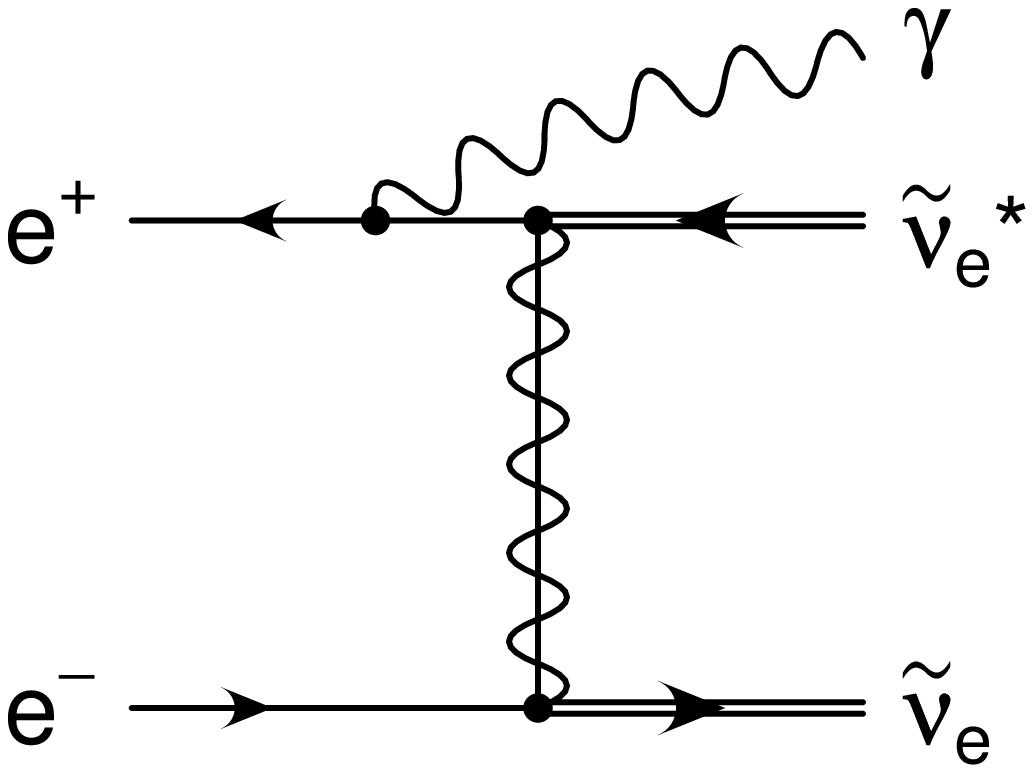, width=5cm}
\hspace{-5mm}
\psfig{figure=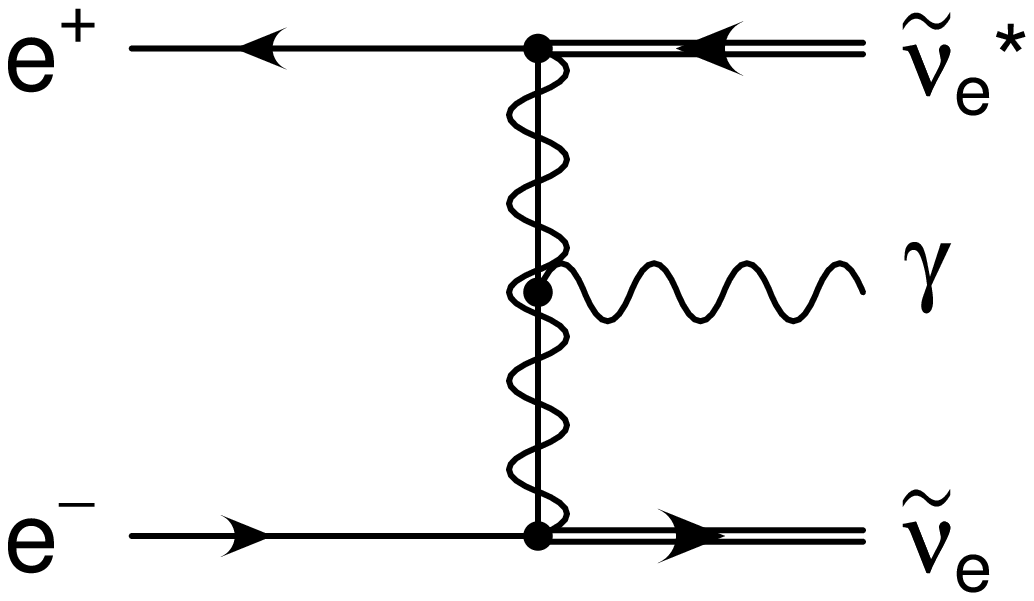, width=5cm}
\hspace{-5mm}\anc\\[-1ex]
(a)\hspace{8cm}(b)\hspace{2.3cm}\anc\\}
\mycaption{Feynman diagrams for real photon emission in sneutrino production,
with (a) contributing to all sneutrino flavors, while (b) only applies for
electron-sneutrino production.
}
\label{fig:realdiag}
\end{figure}
The resulting UV-finite amplitude still contains infrared divergences from
virtual photon exchange, which are absorbed by adding the contributions from
real photon radiation to the cross-section. Since the final-state sneutrinos are
neutral, photon radiation can only occur in the initial state and from internal
charged particle lines, see Fig.~\ref{fig:realdiag}.

In the case of $\sn_\mu$ and $\sn_\tau$ pair production, the analysis is simple
since the virtual
and real (initial-state) QED corrections form a gauge-invariant subset
independent from the other virtual corrections. The QED contributions arising
from virtual and soft real photon radiation are proportional to the Born
cross-section,
\begin{equation}
\begin{aligned}
{\rm d} \sigma_{\rm virt+soft}[e^+ e^- \to \snl \, \snl^*] 
&= {\rm d} \sigma_{\rm Born}[e^+ e^- \to \snl \, \snl^*] 
\times \delta_{\rm virt+soft}^l \\
\delta_{\rm virt+soft}^l = \frac{\alpha}{\pi} &\left [ \log \frac{4(\Delta E)^2}{s}
\left( \log \frac{s}{\me^2} -1 \right) + \frac{3}{2} \, \log \frac{s}{\me^2} -2
+\frac{\pi^2}{3} \right ]
\end{aligned} \qquad
[l \neq e]. \label{eq:qed}
\end{equation}
The dependence on the cut-off $\Delta E$ for the soft-photon energy is
removed when the radiation of hard photons is added to the cross-section.

The loop calculation to $\sne$ pair production is considerably more complex due
to the additional t-channel chargino exchange mechanism. Besides a large set of
additional diagrams, the diagrams involving gauge
bosons, Higgs bosons, gauginos and higgsinos cannot be separated from the QED
loops in a gauge-invariant and UV finite way anymore.
The photonic corrections to the electron-sneutrino-chargino
Yukawa vertex becomes UV finite only after being supplemented by
the corresponding photino loop-diagram. Since the photino is not a mass
eigenstate, this amplitude is intrinsically linked to the remaining degrees of
freedom in the gauge sector. Thus only the total set of gauge boson/Higgs
and gaugino/higgsino electroweak diagrams is gauge invariant. 

Nevertheless, the size of the genuinely process-specific corrections can be
characterized  by the following definition. Subtracting,
from the virtual and soft corrections,
the leading logarithmic
terms $\propto \log (\Delta E)^2/s$ and $\propto \log s/\me^2$, 
which are generated by
infrared/collinear photon emission plus virtual photon
corrections, the remaining corrections are gauge invariant and free of
logarithmically enhanced terms.

\begin{figure}[p]
(a)\\[-1em]
\anc\hfill
\epsfig{file=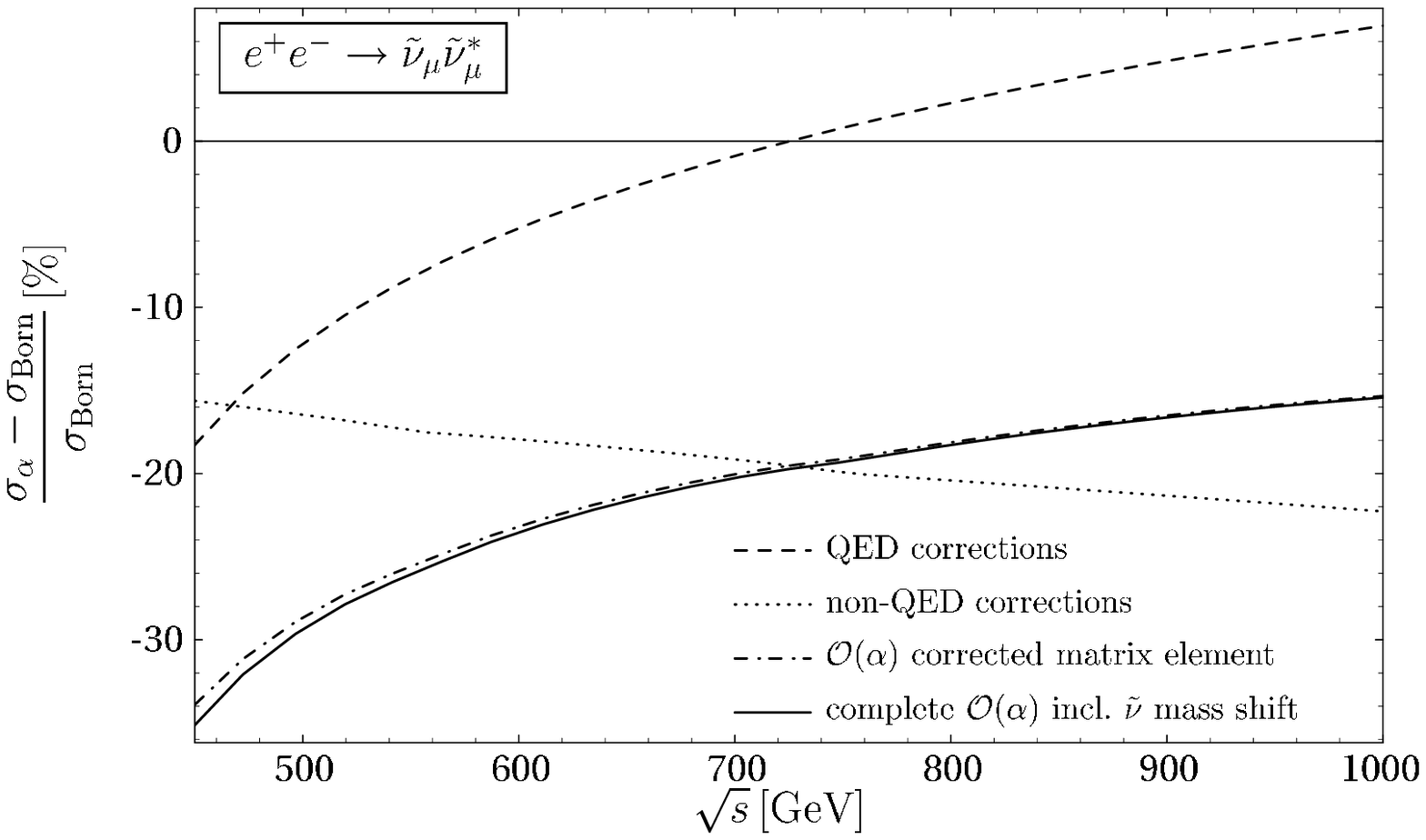,height=3.6in}\\[.7em]
(b)\\[-1em]
\anc\hfill
\epsfig{file=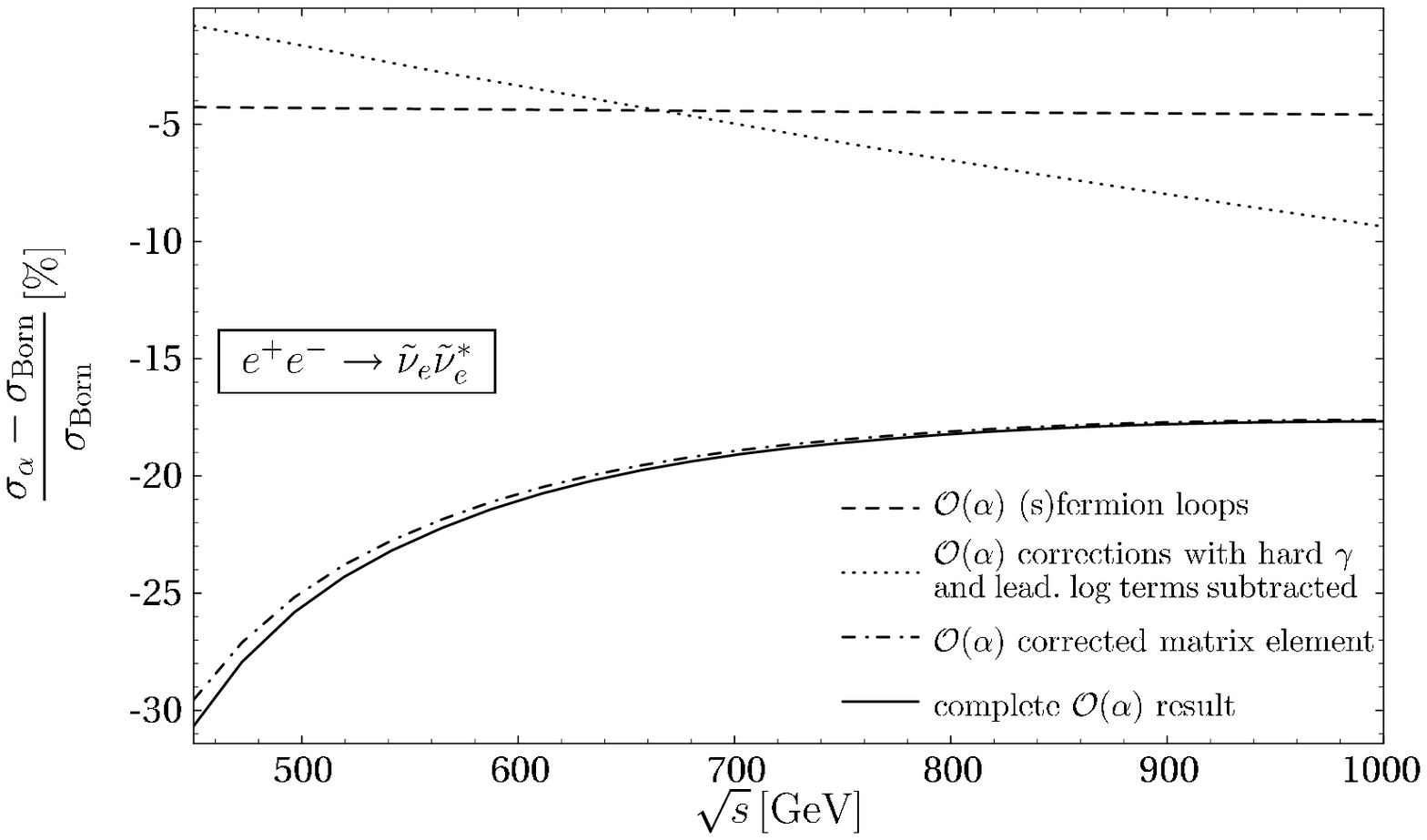,height=3.6in}\\[-1.9em]
\anc
\mycaption{Electroweak corrections to the cross-sections (a) 
for $e^+e^- \to \sn_\mu \sn_\mu^*$ and 
(b) for $e^+e^- \to \sne \sne^*$, relative to the improved Born
cross-section. Besides the full $\OO(\alpha)$ result, contributions from
different subsets of diagrams are shown,
in particular the genuinely process-specific
corrections defined by subtracting hard photon
radiation and leading-log soft and virtual photon effects from the overall $\OO(\alpha)$
corrections. Input parameters taken from 
the SPS1a scenario.}
\label{fig:sn_total}
\end{figure}

The overall next-to-leading order corrections to the total cross-section for 
sneutrino pair production are presented in Fig.~\ref{fig:sn_total}. The
parameters of the Snowmass reference point SPS1a have been adopted again to
illustrate the final results. The corrections are normalized to the Born
cross-section, defined with the running electromagnetic coupling.

In addition to the full corrections, the result for $\sn_\mu$ pair production
is divided into QED corrections and the remaining contributions from massive
virtual loops. Also shown separately is the effect of the one-loop shift to the
sneutrino mass in eq.~\eqref{eq:snshift} in addition to the radiative
corrections to the matrix element ${\cal M}_{(1)}$ in \eqref{eq:snexp}. 
The sneutrino mass shift does not only modify the next-to-leading order
matrix element according to \eqref{eq:snexp}, but is also crucial for defining
the proper sneutrino mass in the physical phase-space kinematics.

For $\sne$ pair production the complete QED subset of the
$\OO(\alpha)$ corrections  cannot be extracted in a gauge-invariant and UV
finite manner. Nevertheless, one can identify a gauge-invariant individual
contribution from closed loops of fermions and/or sfermions, as
shown in Fig.~\ref{fig:sn_total}~(b).
In addition, the dominant QED corrections can be characterized
by the hard photon radiation and the leading logarithmic soft and virtual
contributions. The remaining non-QED logarithmically regularized corrections are
also depicted in Fig.~\ref{fig:sn_total}~(b).

Both for $\sn_\mu$ and $\sne$ pair production, the contribution of the
sneutrino mass shift is relatively small. The total next-to-leading order
corrections, however, are large, between 20 and 30\%. As can be inferred from
Fig.~\ref{fig:sn_total}~(a), these large effects do not only originate from the
QED corrections, which contain large collinear logarithms $\propto \log
s/\me^2$, see eq.~\eqref{eq:qed}. In fact, the massive non-photonic virtual loops
also generate corrections of about 10\% or more.

In the following, the influence of the corrections induced through
the supersymmetry sector itself shall be studied in some detail.
\begin{figure}[p]
{\raggedright (a)\\[-1em]}
\centering{
\epsfig{file=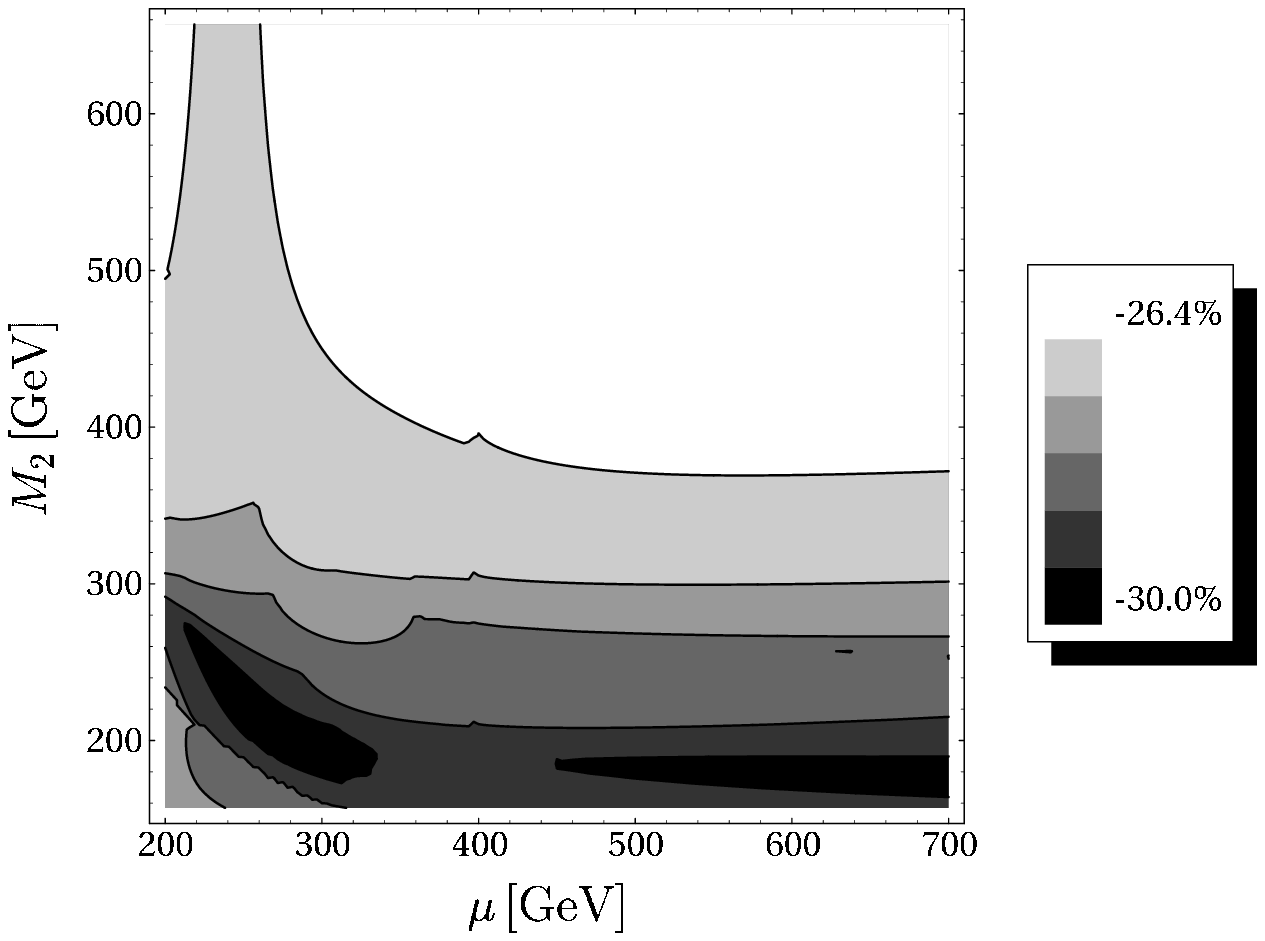,width=5in}
}\\
{\raggedright (b)\\[-1em]}
\centering{
\epsfig{file=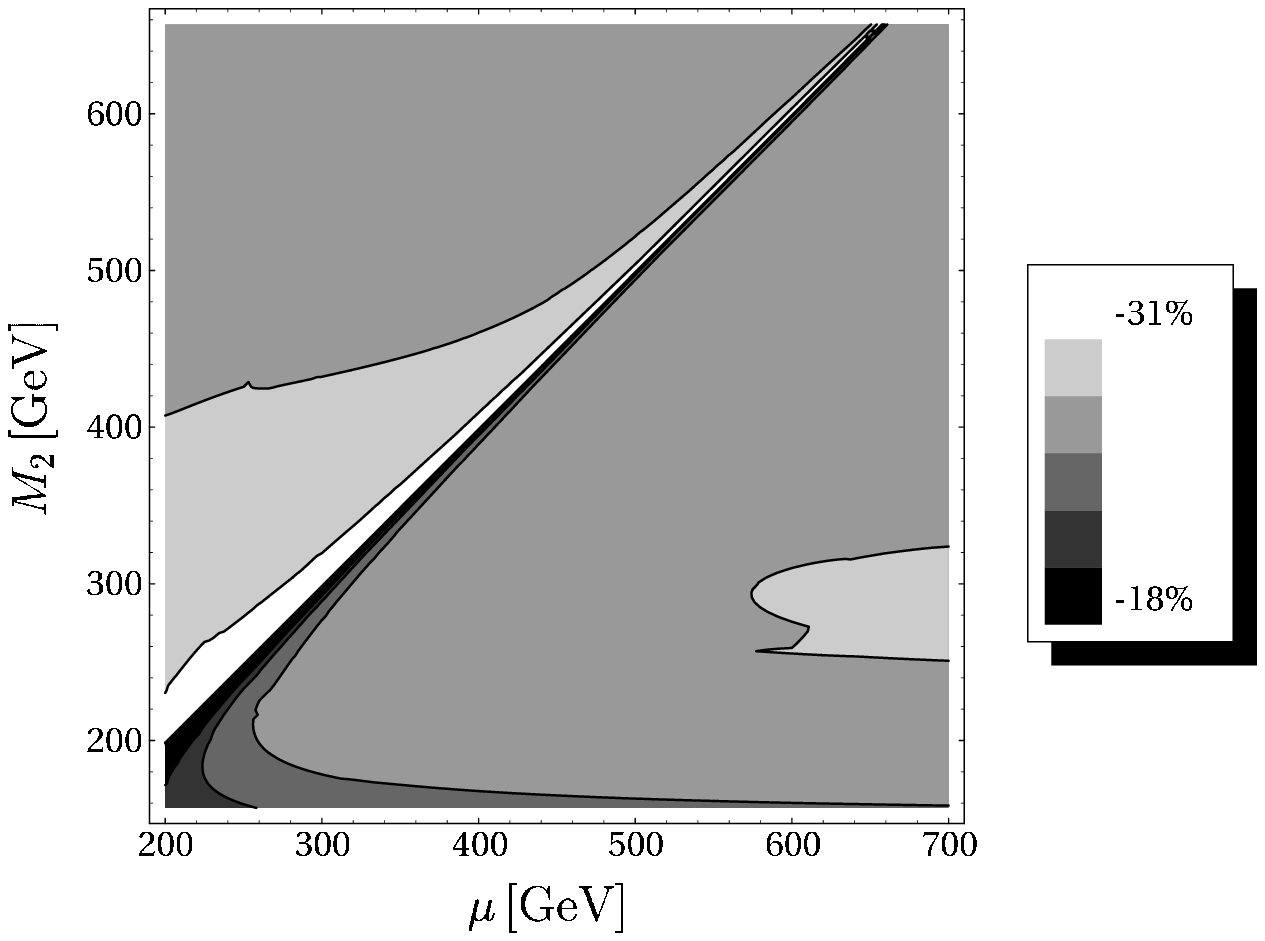,width=5in}
}\\
\vspace{-2ex}
\mycaption{Dependence of the relative one-loop corrections 
$(\sigma_\alpha - \sigma_{\rm Born})/\sigma_{\rm Born}$ to $\sn_\mu \sn_\mu^*$ 
production (a) and
$\sne \sne^*$ production (b) on the gaugino parameters $M_2$ and $\mu$
for $\sqrt{s} = 500$ GeV.
The values of the other parameters are taken from the SPS1a scenario.
}
\label{fig:mum2}
\end{figure}
In Figs.~\ref{fig:mum2}~(a) and (b), the effect of the parameters in the
electroweak gaugino sector on the one-loop corrections is exemplified in regard
to the parameters $M_2$ and $\mu$. The
effects are significant both for $\sn_\mu$ and $\sne$ pair production, with
variations in the cross-section of several per-cent, up to about 10\%. The
abrupt changes along the diagonal $M_2 = \mu$ for $\sne$ pair production in
Fig.~\ref{fig:mum2}~(b) are a consequence of the level crossings between the
$\cha^\pm_i$ states.

As outlined above, large mass splittings between the SUSY sfermions and the
corresponding SM fermions generate large non-decoupling corrections to the
effective Yukawa couplings and thus to $\sne\sne^*$ production. This can be
illustrated by comparing the squark loop effect on the $\sn_\mu$ pair
cross-sections, where supersymmetric Yukawa interactions are absent at the Born
level, with the $\sne$ pair cross-section mediated by t-channel chargino
exchange. 

\begin{figure}[p]
(a)\\[-1em]
\anc\hfill
\epsfig{file=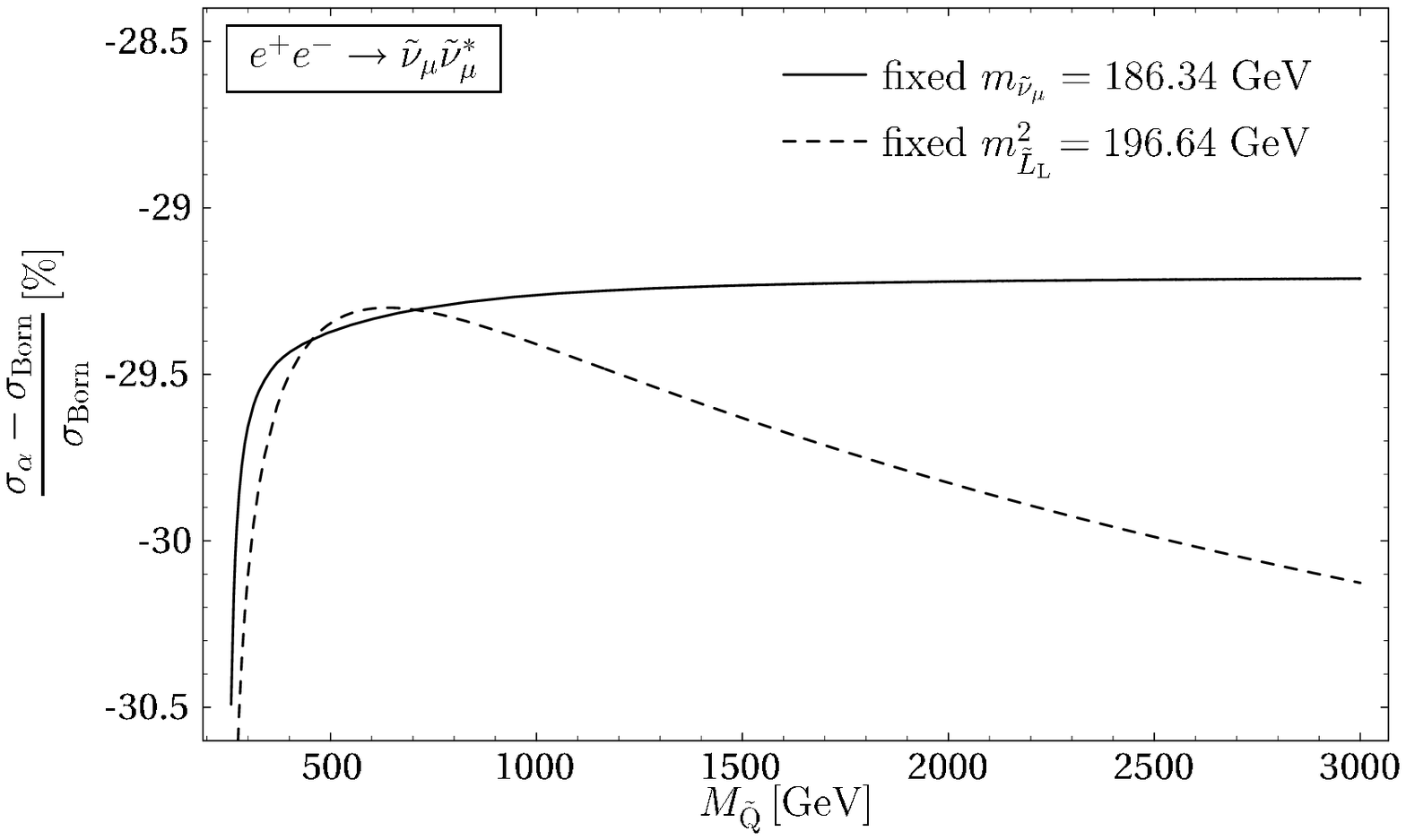,height=3.6in}\\[0.45em]
(b)\\[-1em]
\anc\hfill
\epsfig{file=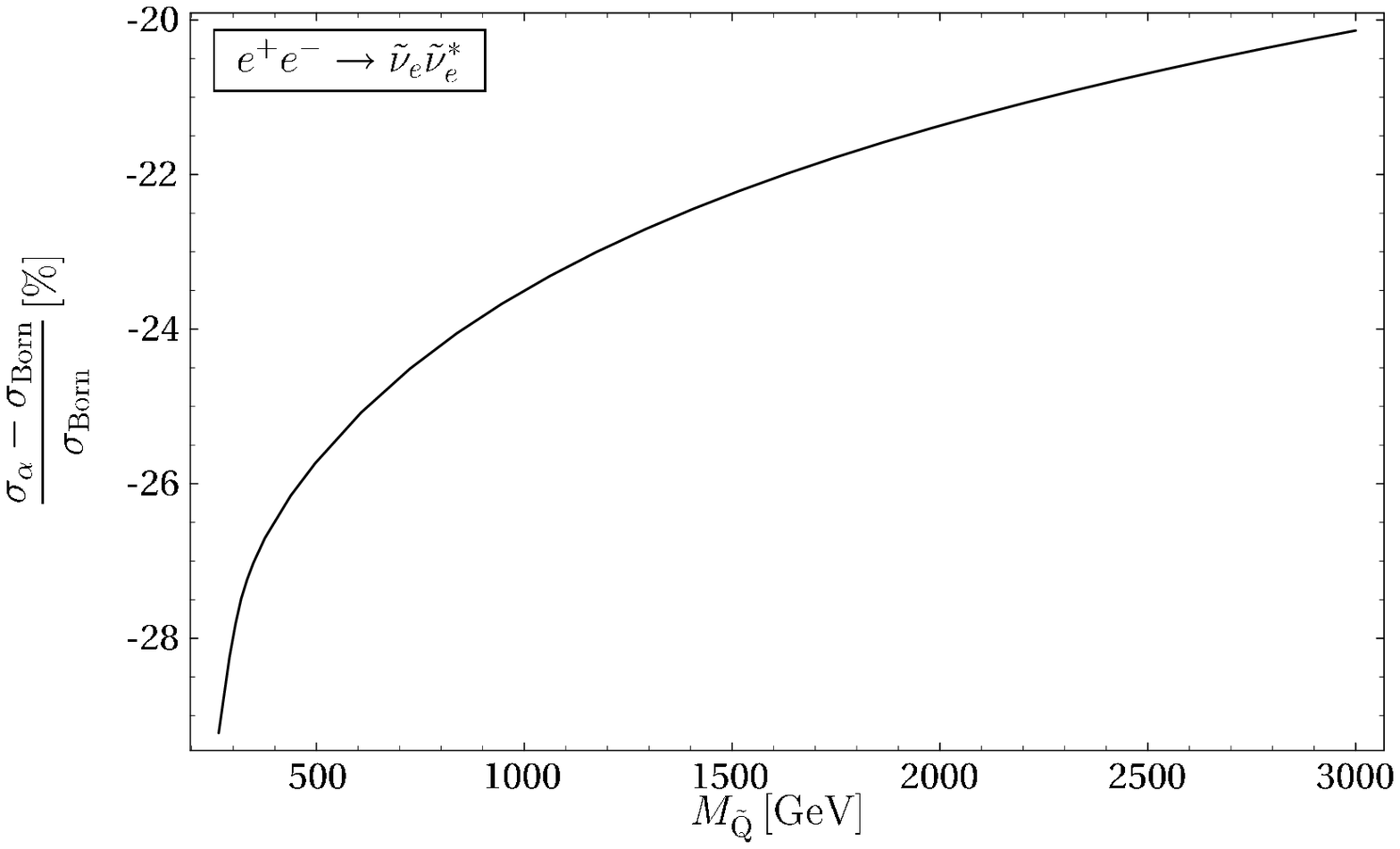,height=3.6in}\\[-2em]
\anc
\mycaption{Dependence of the one-loop corrections on the soft-breaking squark
mass parameter $M_{\rm\tilde{Q}}$ [assumed to be universal for all squarks] for
$\sn_\mu \sn_\mu^*$ production (a) and $\sne \sne^*$ production (b). 
In panel (a) the dashed line corresponds to a fixed L-smuon soft
breaking parameter $m_{\tilde{L}_{\LL}}^2 = 196.64 \gev$ \cite{spsval} as
input, 
while for the full line the physical sneutrino mass $\msn{\mu} = 186.34 \gev$ 
is kept fixed. The values of the other parameters are taken from the SPS1a
scenario, and the cms energy is set to $\sqrt{s} = 500$ GeV.
}
\label{fig:sn_msq}
\end{figure}

Beyond the low-energy region, which is affected by various
thresholds, the radiative loop contributions are expected to rise proportional
to the logarithm $\log M_{\rm\tilde{Q}}$ of the squark masses for $\sne$
production, while approaching a plateau for $\sn_\mu$ production.
This is borne out in the evolution of the full lines in 
Fig.~\ref{fig:sn_msq}~(a) and (b).%
These curves are generated for fixed physical masses of $\sn_\mu$ and $\sne$.
However, as demonstrated by the dashed line in Fig.~\ref{fig:sn_msq}~(a), a small
logarithmic effect is induced also in $\sn_\mu$ pair production if the soft
breaking parameter $m_{\tilde{L}_{\LL}}$ is kept fixed. In this scheme the
physical $\sn_\mu$ mass is shifted by the one-loop corrections in 
eq.~\eqref{eq:snshift}, effecting non-decoupling contributions logarithmic 
in $M_{\rm\tilde{Q}}$

On the other hand, the squark loop contributions to $\sne\sne^*$ production
generate relatively large effects in the effective supersymmetric Yukawa
vertices, with variations of several per-cent for squark masses of $\OO$(1
TeV), cf. Fig.~\ref{fig:sn_msq}~(b). This significantly different behavior
could be used to constrain the parameters of a very heavy squark sector, that
escapes direct detection, through the precise measurement of the $\sne\sne^*$
cross-section, as advocated in Ref.~\cite{Nojiri:97}.

\subsection{The identity of Yukawa and gauge couplings}

The analysis of electron-sneutrino production provides an important testing
ground for the identity of electroweak SUSY Yukawa couplings and the
corresponding SM gauge couplings. This identity can also be tested in
neutralino \cite{ckmz} and selectron \cite{slep} pair production, which involve
electron-selectron-neutralino ($e\se\neu$) Yukawa interactions. While these
channels are sensitive to a mixture of SU(2) and U(1) SUSY Yukawa couplings,
corresponding to the wino and bino components of the neutralinos, there is no
U(1) component contributing to sneutrino pair production, thus allowing direct
access to the SU(2) ($e\sne\tilde{W}$) Yukawa coupling. 
Moreover, the chargino mixing structure is
substantially simpler than the neutralino mixing, thus simplifying the analysis
for $\sne$ pair production compared to $\se$ pair production. This is analogous
to probing the SU(2) Yukawa coupling in chargino pair production, which
provides a complementary experimental method \cite{ckmz}.

In our analysis, we will assume that the masses and mixing parameters of the
charginos can be determined independently from chargino pair production, and
appropriate estimates of the errors for these quantities are included.
Essentially, only the gaugino/higgsino mass parameters $M_2$ and $\mu$ of the
chargino system are important in the sneutrino analysis. The influence of
$\tan\beta$ on the sneutrino cross-section is rather weak, so that knowledge of
$\tan\beta$ to within a factor of two is sufficient. In this case, the MSSM
parameters $M_2$ and $\mu$ can be determined from precision measurements of the
lighter chargino $\chi^\pm_1$ and neutralino $\neu_{1,2}$ masses.

As elaborated in the previous section, the one-loop corrections to the
production cross-section are sizeable and they 
need to be taken into account for any
precision measurement. In principle this will introduce a dependence of the
cross-section prediction on {\it all} MSSM parameters entering in the loops,
including their measurement errors. However, the results can be calculated for
ideal values of the parameters of the squark, charged slepton, Higgs and
neutralino sectors, since their errors would affect the errors in the Yukawa
couplings only to second order. For the parameters present already in the Born
approximation, {\it i.e.} the sneutrino and chargino parameters, errors are
taken into account properly. Iterative procedures may later be employed for the
next-level improvements.

Besides the chargino parameters, the cross-section depends strongly on the
sneutrino masses, which can be extracted from a threshold scan, see
eq.~\eqref{eq:thrres}. The chargino parameters are extracted from mass
measurements of the light chargino and neutralino states, assuming the
following errors,  $\delta \mneu{1} = 50 \mev$, $\delta \mneu{2} = 80 \mev$,
$\delta \mcha{2} = 3000 \mev$, which are based on a coherent analysis of LHC
and LC mass measurements \cite{lhclc}.

The $\sne\sne^*$ production cross-section is computed using the beam
polarizations and cuts introduced in Section~\ref{fsana}, and including
beamstrahlung and ISR effects. Besides increasing the statistics, the
polarization of both the $e^-$ and $e^+$ beams is a tool for testing the chiral
quantum numbers of the sneutrinos. In fact, in the SUSY Yukawa interactions,
$\sne$ only couples to left-handed electrons, while $\sne^*$ only couples to
right-handed positrons. It is assumed that the polarization degree of the
incoming electron/positron beams can be determined with an error of 1\%
\cite{pola}. Detector effects are approximated by assigning a simple global
acceptance factor of $\epsilon_{\rm det} = 50$\%. The numerical analysis in the
SPS1a scenario is based on 500 fb$^{-1}$ data accumulated in $e^+e^-$
collisions at $\sqrt{s} = 500$ GeV.

Taking into account all the constraints and error sources mentioned above, 
the following 1$\sigma$ error for the extraction of the SU(2) 
Yukawa coupling from the cross-section measurement is obtained:
\begin{equation}
\delta \hat{g}/\hat{g} \approx 5\%, \label{eq:yukres}
\end{equation}
where $\hat{g}$ is the SU(2) $e^\pm\sne^{(*)}\cha^\mp_i$ Yukawa coupling.

The precision does not reach the expected sensitivities for the extraction of
the electroweak Yukawa couplings from selectron production
\cite{susyid,superoblique2,slep}, but the sneutrino channel has a less
model-dependent base. The main source of uncertainty contributing to the error
\eqref{eq:yukres} is the sneutrino mass measurement, see eq.~\eqref{eq:thrres},
which due to the dominant invisible decay channel of the sneutrinos is less
precise than for selectrons. The relatively large error is therefore a specific
feature of the SPS1a scenario and the particular sneutrino decay spectrum. In
other scenarios, the determination of the SUSY Yukawa coupling $\hat{g}$ from
the sneutrino cross-section can improve considerably, see {\it e.g.}
Ref.~\cite{Nojiri:97}.

In summary, it turns out that the analysis of electron-sneutrino pair
production at $e^+e^-$ colliders can test the fundamental identity of
Yukawa and gauge couplings in supersymmetric theories at the level of a few
per-cent.


\section{Conclusions}
\label{concl}

This report extends previous work in Ref.~\cite{slep} on precision studies of
scalar leptons at future linear colliders, by analyzing the sector of neutral
scalar leptons. We have studied the pair production of sneutrinos in $e^+e^-$
collisions, discussing the theoretical basis and drawing the phenomenological
consequences for the determination of the masses and couplings of these
particles.

The clean environment of high-energy lepton colliders operating with polarized
beams at high luminosity allows one to measure the properties of SUSY
particles with high precision. From a scan of the production cross-section near
threshold, the \uline{mass of the electron-sneutrino
$m_{\sne}$} can be determined with
unparalleled precision due to the characteristic rise of the excitation curve
with the third power in the sparticle velocity.

In order to provide a reliable theoretical description of the threshold
cross-section, it is necessary to include non-zero width effects in a
consistent way. The analysis is complicated by the fact that in typical
mSUGRA-inspired MSSM scenarios, the sneutrino dominantly decays invisibly.
Nevertheless, it could be shown that a significant signal can be extracted by
combining one visible with one invisible decay mode 
of the sneutrino pair, and that
backgrounds both from Standard Model and supersymmetric sources can be
sufficiently reduced.

Using these elements, a phenomenological analysis of slepton masses in
threshold scans was performed. 
Despite the impact of the relative low signal rate and remaining sub-dominant
backgrounds, a precision of better than the per-cent level for the
electron-sneutrino mass determination can be reached. The threshold fit can also
be used to experimentally constrain the total decay width of the
electron-sneutrino, but the accuracy would not suffice to establish significant
experimental evidence for a non-zero sneutrino width.

The measurement of the electron-sneutrino cross-section can also be exploited
to test the fundamental equality of the \uline{Yukawa couplings
$\hat{g}(e\sne\tilde{W})$} of the electron, the sneutrino and the SU(2)
gauginos, and the corresponding gauge couplings  $g(e \nu_e W)$ of the 
electrons and electron-neutrinos 
to the $W$ gauge bosons. Due to the chargino t-channel exchange, 
electron-sneutrino
pair production is sensitive to the SUSY SU(2) Yukawa coupling. Based on a
careful analysis of statistical errors and systematic uncertainties, we found
that the SU(2) Yukawa coupling can be extracted at the few per-cent level or
better from total cross-section measurements in the high-energy continuum.

Accordingly, the theoretical predictions for the cross-sections need to be
calculated to the per-cent level to match this expected experimental accuracy.
To this end, the complete next-to-leading order SUSY electroweak corrections
have been calculated for the production of on-shell sneutrino pairs in the
continuum. These one-loop corrections can be relatively large, up to the order
of 20--30\%, with genuine SUSY loop contributions leading to sizeable effects
of a few per-cent. Thus the sneutrino pair production cross-sections are
significantly affected by parameters of other SUSY sectors, which must be
introduced in systematic iterative procedures.

We have implemented the results into computer programs, that are available on
the web at \texttt{http://theory.fnal.gov/people/\linebreak[0]afreitas/}.
[Technical information on installing and running the programs are given at this
web site.]

A self-consistent analysis of the SUSY sector at the expected experimental
precision will eventually have to take into account the impact of the
parameters of all SUSY sectors in the theoretical predictions at the loop
level. Accordingly, a coherent study of many different channels in
parallel---sleptons, charginos/neutralinos and squarks/gluinos---has to be
performed \cite{sfitterino}. Such a comprehensive study is currently pursued in
the context of the SPA Project~\cite{spa}, and will ultimately establish the
access to the fundamental supersymmetric theory and its microscopic breaking
mechanism at potentially high scales.

\section*{Acknowledgements}

We benefited from very helpful communication with H.~U.~Martyn and U.~Nauenberg
on experimental sneutrino analyses at $e^+e^-$ linear colliders. We are also
very grateful to G.~A.~Blair for the careful reading of the manuscript.


\end{document}